\documentclass[aps,pre,reprint,nofootinbib]{revtex4-1}

\usepackage[utf8]{inputenc}
\usepackage{amsmath}
\usepackage{amsfonts}
\usepackage{amssymb}
\usepackage{amsthm}
\usepackage{amsopn}
\usepackage{upgreek}
\usepackage{bm}
\usepackage{epsfig}
\usepackage{units}
\usepackage{enumitem}
\usepackage{nicefrac}

\renewcommand{\eqref}[1]{Eq.~(\ref{#1})}

\newcommand{\figref}[1]{Figure~\ref{#1}}

\newcommand{\refcite}[1]{Ref.~\onlinecite{#1}}

\newcommand{\kB}{k_{\text{B}}}

\newcommand{\addressharvard}{Department of Chemistry and Chemical Biology, Harvard University, 12 Oxford Street, Cambridge, MA 02138, USA}

\begin{document}

\title{Accurate protein-folding transition-path statistics from a simple free-energy landscape}
\author{William M.~Jacobs}
\affiliation{\addressharvard}
\author{Eugene I.~Shakhnovich}
\affiliation{\addressharvard}
\date{\today}

\begin{abstract}
A central goal of protein-folding theory is to predict the stochastic dynamics of transition paths --- the rare trajectories that transit between the folded and unfolded ensembles --- using only thermodynamic information, such as a low-dimensional equilibrium free-energy landscape.
However, commonly used one-dimensional landscapes typically fall short of this aim, because an empirical coordinate-dependent diffusion coefficient has to be fit to transition-path trajectory data in order to reproduce the transition-path dynamics.
We show that an alternative, first-principles free-energy landscape predicts transition-path statistics that agree well with simulations and single-molecule experiments without requiring dynamical data as an input.
This `topological configuration' model assumes that distinct, native-like substructures assemble on a timescale that is slower than native-contact formation but faster than the folding of the entire protein.
Using only equilibrium simulation data to determine the free energies of these coarse-grained intermediate states, we predict a broad distribution of transition-path transit times that agrees well with the transition-path durations observed in simulations.
We further show that both the distribution of finite-time displacements on a one-dimensional order parameter and the ensemble of transition-path trajectories generated by the model are consistent with the simulated transition paths.
These results indicate that a landscape based on transient folding intermediates, which are often hidden by one-dimensional projections, can form the basis of a predictive model of protein-folding transition-path dynamics.
\end{abstract}

\maketitle

\section{Introduction}

In studies of complex molecular systems, free-energy landscapes provide a tractable and intuitive framework for predicting rare events.
Free-energy landscapes are low-dimensional projections that describe the equilibrium distribution of molecular configurations with respect to a small number of collective variables.
However, when appropriately defined, these landscapes can also be used to predict dynamical properties of equilibrium or near-equilibrium stochastic trajectories, such as the relative rates of transitions between macrostates.
This feature, combined with the fact that a variety of computational techniques have been developed for efficiently calculating free energies without directly simulating rare events~\cite{frenkel2001understanding}, makes free-energy landscapes useful for rationalizing reaction and phase-transformation mechanisms in complex systems.
Consequently, theories based on free-energy landscapes are widely applied to problems in classical~\cite{gibbs1878equilibrium,rein1996numerical,de2003principles} and non-classical~\cite{ten1997enhancement,jacobs2015rational} nucleation, phase separation~\cite{frenkel1999entropy,poon2000kinetics}, and protein folding~\cite{abkevich1994free,shakhnovich2006protein,gfeller2007complex,thirumalai2010theoretical}.

A particularly important problem in protein folding is the prediction of transition paths between the unfolded and folded ensembles~\cite{abkevich1994specific,shakhnovich1997theoretical,chung2012single,chung2013single,chung2013measuring,truex2015testing}.
These \mbox{trajectories are} both rare and, at an atomistic level, extremely heterogeneous, making this problem ideal for landscape-based theories.
One of the most widely adopted approaches is to model folding as a diffusion process on a smooth, one-dimensional free-energy landscape~\cite{sali1994how,bryngelson1995funnels,onuchic2004theory,berezhkovskii2005one,best2006diffusive,zhang2007transition,best2011diffusion}.
Nevertheless, even when using a \mbox{good reaction coordinate},\footnote{In this context, a good reaction coordinate is one that not only distinguishes the unfolded and folded states, but also takes a single value for all configurations visited on transition paths that have an equal probability of reaching either of these macrostates.} one-dimensional landscapes typically have to be corrected empirically to reproduce the dynamical properties of the actual stochastic folding trajectories~\cite{hummer2005position,best2011diffusion}.
This correction can be achieved by introducing a position-dependent diffusion coefficient~\cite{berezhkovskii2011time}, since the gradient of the free-energy landscape itself is not sufficient to predict the relative rates of molecular motions on the one-dimensional reaction coordinate.
The key limitation of this approach is that the transition-path trajectories that we wish to predict are required as input, either to determine the coordinate-dependence of the diffusion coefficient~\cite{hummer2005position,best2005reaction,hinczewski2010diffusivity,mugnai2015extracting} or to find a projection for which the apparent diffusive behavior is coordinate-independent~\cite{best2010coordinate,tiwary2017predicting}.
It is also unclear whether a single optimized one-dimensional coordinate can always be found for large proteins, which may have more complicated or parallel folding pathways~\cite{du1998transition}.
Furthermore, recent single-molecule measurements of folding transition paths~\cite{chung2012single,neupane2016direct} have provided experimental evidence of the shortcomings of one-dimensional landscapes, as the folding free-energy barrier inferred by applying a one-dimensional diffusion model to measured transit-time distributions is often inconsistent with the landscape determined directly in the same experiments~\cite{makarov2017reconciling,satija2017transition}.

In contrast, an optimal free-energy landscape is one that is capable of predicting the statistical properties of stochastic transition paths without requiring additional, empirical kinetic information.
To address this problem, we recently proposed an alternative, first-principles approach~\cite{jacobs2016structure} for constructing structure-based free-energy landscapes to describe protein-folding transition paths.
Based on an analysis of a native-centric `Ising-like' model~\cite{munoz1999simple,alm1999prediction,galzitskaya1999theoretical}, we postulated that the key events along transition paths coincide with the formation of native-like loops in the polymer backbone~\cite{frenkel2016folding}.
We therefore devised a coarse-graining procedure in which microstates sharing the same set of native-like loops, but different sets of native contacts, are grouped into the same `topological configuration.'
Because the loss of entropy due to loop closure is not compensated until multiple stabilizing native contacts are formed, we further postulated that these topological configurations are, in general, separated by free-energy barriers, leading to a separation of timescales between the formation of individual native contacts and the assembly of topological configurations.
Consistent with this prediction, we found that topological configurations interconvert on much slower timescales than individual native contacts in atomistic simulations and that these slower transitions follow roughly Markovian dynamics~\cite{jacobs2016structure}.

In this paper, we show that the topological configuration model accurately describes the stochastic dynamics of transition-path trajectories.
By estimating the free energies of the predicted topological configurations using equilibrium all-atom simulation snapshots~\cite{piana2013atomic}, we apply this model to generate an ensemble of transition paths in terms of transitions between coarse-grained, partially folded states.
First, we show that the distribution of transit times predicted by this approach is much closer to the distribution of simulated transit times than that predicted by a model of diffusion on a smooth one-dimensional landscape.
Second, we demonstrate that the distribution of simulated finite-time displacements on a one-dimensional order parameter can be rationalized by the topological configuration model.
Lastly, we use a hidden Markov framework to show that the predicted separation of timescales generates an ensemble of transition paths that is consistent with the simulated folding trajectories.
Overall, these results indicate that a free-energy landscape defined on the basis of transient, native-like intermediates can predict protein-folding transition paths without requiring \textit{post hoc} corrections to the transition-path dynamics.

\section{Theory}

\subsection{Definition of a topological configuration}

\begin{figure}
  \includegraphics{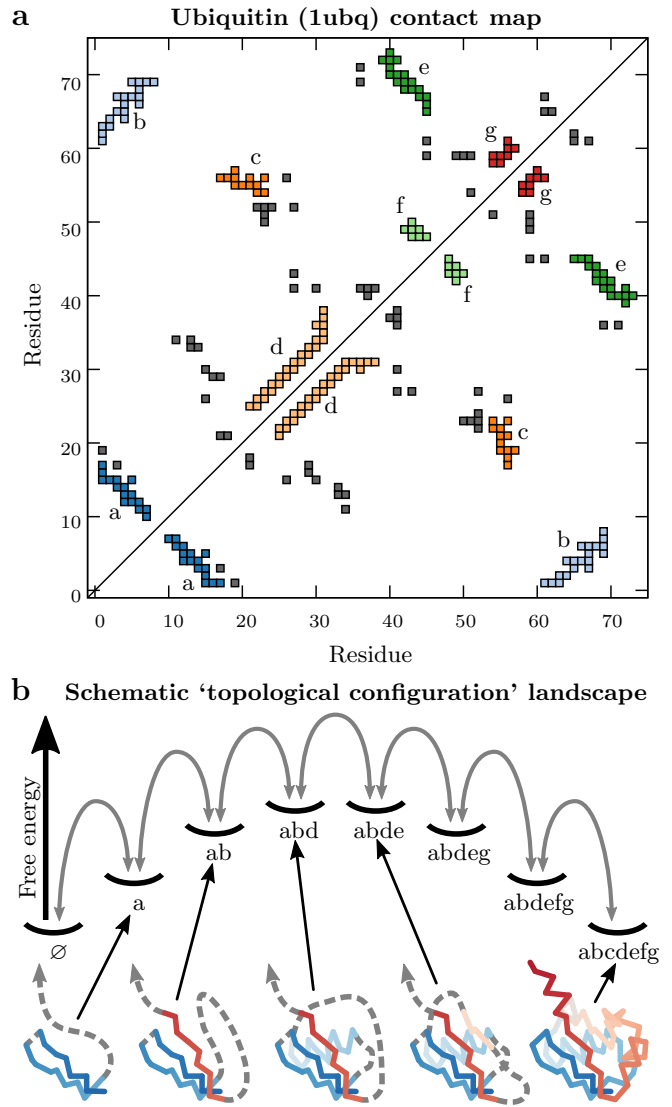}
  \caption{Definition of topological configurations for the protein Ubiquitin.  (a)~A map of the residue--residue native contacts determined from a crystal structure (PDB ID: 1ubq).  Substructures, comprising sets of at least six adjacent native contacts, are colored and labeled.  Native contacts that are not part of any substructure are shown in gray.  (b)~A schematic topological configuration free-energy landscape showing a single pathway between the completely unfolded, `$\varnothing$,' and native, `abcdefg,' configurations.  These configurations are separated by free-energy barriers and thus interconvert on a slower timescale than the formation of individual native contacts.  Selected configurations are illustrated below, where stretches of disordered residues are indicated by dashed lines.}
  \label{fig:schematic}
\end{figure}

The central principle of the topological configuration model~\cite{jacobs2016structure} is a separation between three different timescales: a relatively fast timescale associated with native-contact formation, a slower timescale on which substantial portions of native structure assemble, and a slowest timescale on which the entire protein folds.
The timescale separation for these processes has been well established experimentally, with measurements reporting single-contact formation on timescales of approximately $10\,\text{ns}$~\cite{thompson1997laser}, native-like loop formation on timescales of $100\,\text{ns}$--$1\,\mu\text{s}$~\cite{lapidus2000measuring}, and the folding of proteins with approximately 100 residues or more on timescales of $100\,\mu\text{s}$ or longer~\cite{garbuzynskiy2013golden}.
It is also well established that the entropy--enthalpy compensation of protein folding is imperfect, meaning that while a small number of native contacts is rarely sufficient to counter the associated loss of configurational entropy completely, the formation of subsequent native contacts tends to be more thermodynamically favorable~\cite{zimm1959theory,dill1993cooperativity}.
This general feature, which is responsible for the overall free-energy barrier that determines the slowest timescale of protein folding, also gives rise to many smaller barriers on folding transition paths.
In particular, the entropic penalty associated with the closure of a single native-like loop typically results in a small yet significant free-energy barrier, which in turn leads to a dynamical timescale that is slower than the average rate of native-contact formation but faster than the rate at which the entire protein folds.

By identifying native loops and considering all permutations of the order in which they can form, we can construct a free-energy landscape that captures this key intermediate timescale.
We previously demonstrated~\cite{jacobs2016structure} how this analysis could be applied to a structure-based, `Ising-like' model based on the pioneering work of Eaton and colleagues~\cite{munoz1999simple,kubelka2008chemical,henry2013comparing}.
In that work, we calculated the free-energy barriers between configurations in the structure-based model to support the predicted separation of timescales.
We also provided indirect evidence of intermediate barriers in an all-atom model by analyzing the dwell times associated with the predicted topological configurations in all-atom simulations.

As in \refcite{jacobs2016structure}, we shall focus on the protein Ubiquitin in order to test the topological configuration model's ability to predict transition-path dynamics.
In the contact map shown in \figref{fig:schematic}a, individual `substructures' comprise native contacts that are adjacent\footnote{Two contacts are adjacent if each residue in the first contact is an immediate neighbor of one of the residues in the second contact; for example $(i,j)$ is adjacent to $(i,j+1)$ and $(i+1,j+1)$.} in the contact map.
Topological configurations are defined as combinations of one or more substructures, including any additional native contacts between the residues that comprise these substructures.
As shown in the schematic \figref{fig:schematic}b, we can then construct a free-energy landscape in the discrete space of topological configurations.
Transitions are allowed between configurations that differ by a single substructure (and thus a single native-like loop closure).
It is important to note that each topological configuration is not a rigid structure but rather an ensemble of microstates, in which different sets of native contacts are present but the same set of substructures (and, consequently, native-like loops) are represented.
Furthermore, due to the predicted separation of timescales, the fluctuations in native-contact formation within a topological configuration are typically much faster than the transitions between configurations.

\subsection{Estimation of topological configuration free energies from Molecular Dynamics simulations}

\begin{figure}
  \includegraphics{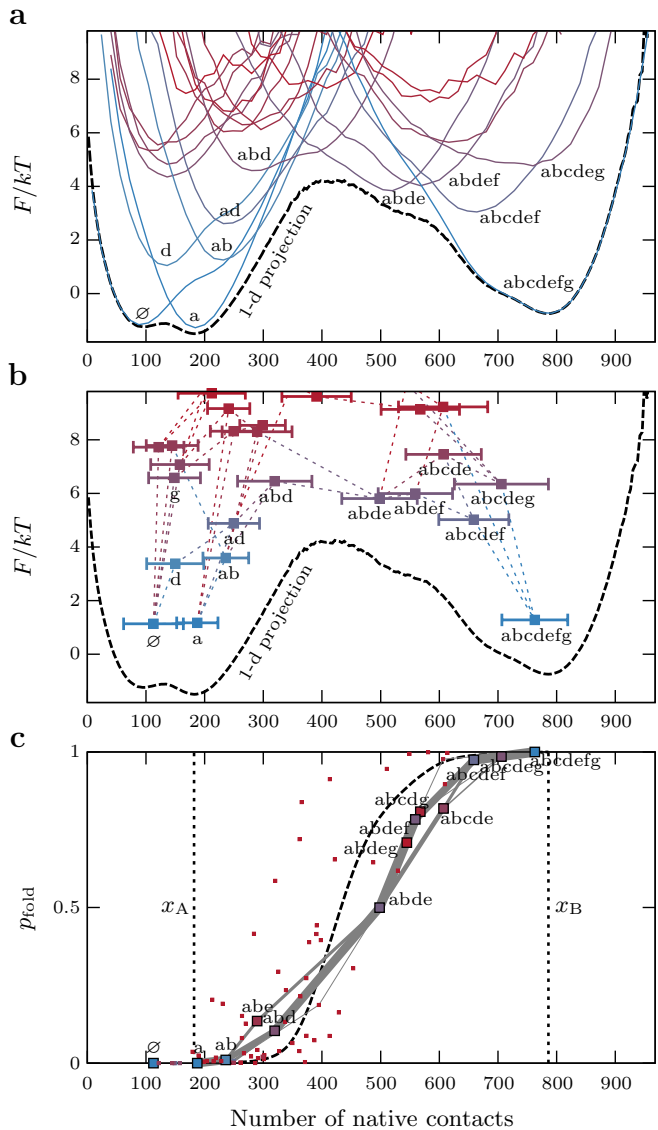}
  \caption{Comparison of the topological configuration and one-dimensional (\mbox{1-d}) free-energy landscapes.  (a)~Projections of the free-energy surfaces onto the number of native contacts.  Selected curves corresponding to distinct topological configurations (solid colored lines) are labeled; the \mbox{1-d} projection $F(x)$, with bin width ${\Delta x = 4}$ native contacts, is shown by the dashed black line.  (b)~The unimodal projections of the topological configurations onto the number of native contacts can be characterized by Gaussian distributions with the indicated means (squares) and standard deviations (error bars).  The dotted lines indicate allowed transitions between configurations that differ by exactly one substructure. (c)~The committors, $p_{\text{fold}}$, associated with each configuration (squares), projected onto the number of native contacts.  Selected configurations are labeled, and the predicted folding fluxes through the network of states are indicated by the widths of the gray lines; fluxes less than 5\% have been omitted.  For comparison, the dashed black line shows $p_{\text{fold}}(x)$ predicted by the \mbox{1-d} landscape, where the vertical dotted lines indicate the boundaries of the transition-path region, $x_{\text{A}}$ and $x_{\text{B}}$.  In all panels, the configurations are colored according to their estimated free energies, ${F_i = -k_{\text{B}}T \log N_i}$, with blue and red indicating low and high free energies, respectively.}
  \label{fig:landscapes}
\end{figure}

In this paper, our goal is to evaluate the predictions of a free-energy landscape constructed solely from thermodynamic data.
For this purpose, we analyzed snapshots from all-atom Molecular Dynamics (MD) simulations~\cite{piana2013atomic}, which were conducted under conditions where a total of ten unbiased, reversible folding and unfolding transition paths of Ubiquitin were observed.
To calculate the free energy associated with each predicted topological configuration, we first classified all simulation snapshots, recorded at ${1\,\text{ns}}$ intervals, according to which substructures are present.
In each snapshot, we found all native residue--residue contacts in which at least one pair of heavy atoms is less than ${4.5\,\text{\AA}}$ apart.
We then considered a substructure to be formed if at least six of its native contacts were present in the largest structured region, i.e., the largest connected component of the graph of native residue--residue contacts in a simulation snapshot.
As demonstrated in \refcite{jacobs2016structure}, this definition prevents contacts with extremely brief lifetimes (on the order of $1\,\text{ns}$) from influencing the identification of substructures.
Free energy differences between pairs of configurations $(i,j)$ were estimated according to the relative frequencies of the classified snapshots,
\begin{equation}
  \Delta F_{ij} = -k_{\text{B}}T \log \left( N_i / N_j \right),
\end{equation}
where $N_i$ is the total number of snapshots assigned to configuration $i$, $k_{\text{B}}$ is the Boltzmann constant, and $T$ is the absolute temperature.
For comparison with theories based on one-dimensional free-energy landscapes, we also calculated the free-energy landscape as a function of the number of native heavy-atom contacts~\cite{best2013native} using a ${4.5\,\text{\AA}}$ cutoff distance,
\begin{equation}
  F(x) = -k_{\text{B}}T \log N(x) + \text{const}.,
\end{equation}
where $N(x)$ is the total number of snapshots in which the number of native contacts falls in the range ${[x-\Delta x/2, x+\Delta x/2)}$.
This one-dimensional free-energy landscape is shown in \figref{fig:landscapes}a, where the bin width $\Delta x$ is taken to be 4 native contacts.

The central object of the topological configuration model is the rate matrix $\bm{T}$ for transitions between topological configurations.
Ideally, one should determine the transition rates either from the free-energy barriers or the mean first passage times between configurations, but accurate calculations of this type are not possible given the available simulation data.
Instead, we simply assumed a symmetric form that enforces detailed balance for the forward and backward rates between configurations $i$~and~$j$,
\begin{equation}
  \label{eq:rates}
  T_{ij} = k_0 \exp\left(-\frac{\Delta F_{ji}}{2 k_{\text{B}} T}\right) \quad i \ne j,
\end{equation}
for pairs of configurations $(i,j)$ that differ by the addition or removal of a single substructure; the diagonal elements of the matrix are then ${T_{ii} = -\sum_{j \ne i} T_{ij}}$.
The prefactor $k_0$ is the same for all transitions and is left as an adjustable parameter that scales all barrier heights between configurations equivalently.
However, due to the assumed separation of timescales, we know that $k_0$ should be slow compared to the average rate of native-contact formation.
Then, given the matrix $\bm{T}$, it is straightforward to calculate the overall folding rate, $k_{\text{fold}}$; the committor associated with each configuration $i$, $p_{\text{fold},i}$; the probability of finding a trajectory in a specific configuration $i$ on a transition path, $m^{\text{AB}}_i$; and the folding fluxes between adjacent states $i$ and $j$, $f^{\text{AB}}_{ij}\!$, using transition-path theory~\cite{metzner2009transition}.
These quantities will be used throughout our analysis.
We shall show that, despite not undertaking detailed calculations of the individual barrier heights, this model produces remarkably accurate transition-path statistics.
Furthermore, alternative choices for the form of the rates given in \eqref{eq:rates}, such as a Metropolis function~\cite{metropolis1953equation}, do not change the qualitative nature of our results.
We also note that all timescales determined directly from the atomistic simulations are accelerated relative to experiments, in part due to the elevated temperature at which the simulations were conducted~\cite{piana2013atomic}.

\subsection{General properties of topological configuration free-energy landscapes}

In addition to a separation of timescales, our previous analysis~\cite{jacobs2016structure} of this model made two general predictions that hold up when comparing with simulation data.
First, the Boltzmann-weighted ensemble of microstates associated with each topological configuration is predicted to be unimodal when projected onto a one-dimensional coordinate.
For example, the free energy as a function of the number of native contacts is unimodal for all topological configurations, as shown by the labeled colored curves in \figref{fig:landscapes}a, suggesting that there are no significant free-energy barriers between microstates within each configuration.
Consequently, it is reasonable to approximate the projection of each configuration onto this order parameter using a Gaussian distribution with the estimated mean, ${\langle x \rangle_i}$, and variance, ${\langle x^2 \rangle_i - \langle x \rangle^2_i}$, where the subscripts indicate averages over all snapshots classified as topological configuration $i$, as shown in \figref{fig:landscapes}b.
This approximation will be used in the discussion of hidden Markov modeling below.

Second, in the case of proteins such as Ubiquitin with little structural symmetry, the free energies of the various topological configurations are relatively heterogeneous, which results in a small number of high-probability transition paths through the network.
By applying transition-path theory~\cite{metzner2009transition} to the rate matrix $\bm{T}$, we calculated $p_{\text{fold}}$ for each configuration and the folding flux between configurations on transition paths.
\figref{fig:landscapes}c shows that there is a nearly one-to-one correspondence between the predicted $p_{\text{fold}}$ and the number of native contacts when we consider only those configurations that are likely to be visited on transition paths, even though this order parameter played no role in the transition-path theory calculations.
This observation is consistent with the fact that the number of native contacts is a good reaction coordinate for identifying the ensemble of transition states, where ${p_{\text{fold}} = \nicefrac{1}{2}}$, from simulated Ubiquitin transition paths~\cite{best2013native}.
However, knowing the location of the transition state on an order parameter is, in general, not sufficient to predict the transition-path dynamics.
In addition, any one-dimensional projection almost invariably hides some of the intermediate free-energy barriers on transition paths that play an important role in the transition-path dynamics~\cite{krivov2004hidden,krivov2006one}.

\section{Statistical analyses}

\subsection{Distribution of transition-path transit times}

As an initial test of the model, we examined the predicted distribution of transit times between the unfolded and folded ensembles.
For two-state proteins, transit times are orders of magnitude smaller than the characteristic waiting time until a folding or unfolding event occurs~\cite{abkevich1994specific,ding2002direct,chung2009experimental,chung2012single}.
Nevertheless, recent advances in single-molecule experiments~\cite{chung2012single,neupane2016direct} have made it possible to measure these brief trajectories.
Although experimental measurements have confirmed that the distribution of transit times has an exponential tail as expected for stochastic barrier-crossing processes, the shape of the distribution generally disagrees with the predictions of one-dimensional landscape theories~\cite{neupane2016direct,makarov2017reconciling,satija2017transition}.
In particular, the measured distributions are typically much broader than expected for a one-dimensional landscape with a harmonic barrier.

\begin{figure}
  \includegraphics{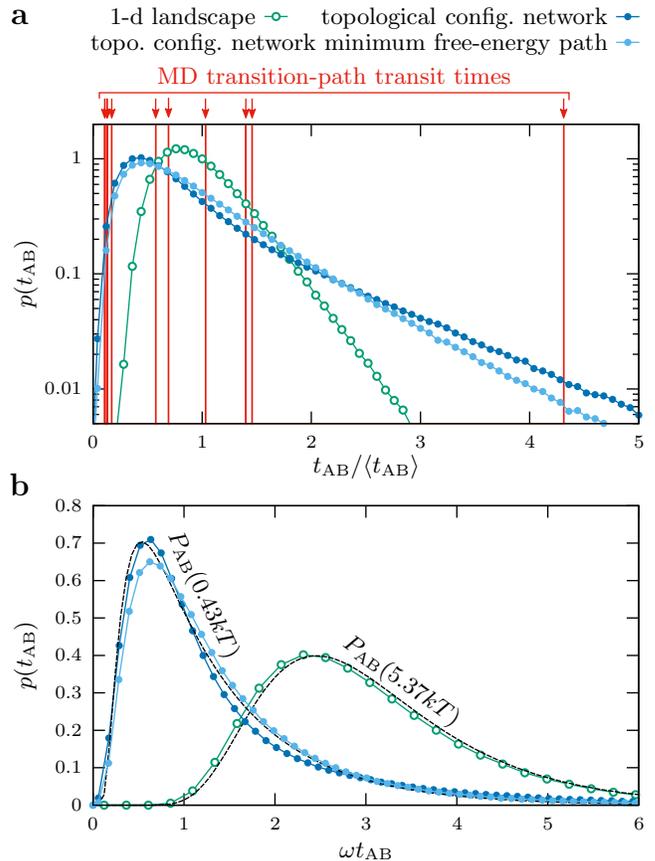}
  \caption{Distributions of transition-path transit times, $p(t_{\text{AB}})$, for the models of Ubiquitin folding shown in \figref{fig:landscapes}.  (a)~The distributions calculated via kinetic Monte Carlo simulations of the \mbox{1-d} native contacts landscape model, the topological configuration model, and the quasi-one-dimensional minimum free-energy path through the topological configuration network.  To compare the shapes of the distributions, all transit times are scaled according to the mean transit time, ${\langle t_{\text{AB}} \rangle}$, for each model.  Also shown are the ten transit times observed in all-atom MD simulations.  (b)~The same three distributions were fit to the theoretical distribution for a harmonic barrier, $P_{\text{AB}}^{\text{harm}}\!$, in order to estimate the decay constant, $\omega^{-1}\!$, of the exponential tail.  The transit-time distribution calculated from the empirical \mbox{1-d} landscape agrees well with the theoretical distribution using the empirical barrier height, ${5.37kT}$, while the distribution calculated from the topological configuration model can only be fit by ${P_{\text{AB}}^{\text{harm}}}$ if we set the shape parameter $\Delta F^\dagger$ equal to a much lower barrier height of ${0.43 kT}$.}
  \label{fig:tp_times}
\end{figure}

To compare the predictions of the one-dimensional \mbox{(1-d)} and topological configuration models, we used kinetic Monte Carlo (kMC) simulations~\cite{gillespie1977exact} to sample transition paths between absorbing states of a rate matrix, $\bm{T}$.
We first calculated the distribution of transit times for the \mbox{1-d} native-contacts landscape shown in \figref{fig:landscapes}a.
To this end, we discretized this landscape between the unfolded and folded free-energy minima into 150 bins (such that ${\Delta x = 4}$ native contacts) and constructed a tri-diagonal transition matrix, $\bm{T^{\text{1-d}}}$.
We assumed the symmetric form
\begin{equation}
  \label{eq:rates1d}
  T^{\text{1-d}}_{(x,x\pm\Delta x)} = k_0 \exp\left[ -\frac{F(x \pm \Delta x) - F(x)}{2k_{\text{B}}T}\right],
\end{equation}
with ${T^{\text{1-d}}_{(x,x)} = -[T^{\text{1-d}}_{(x,x-\Delta x)\!} + T^{\text{1-d}}_{(x,x+\Delta x)}]}$.
Because the transit times are inversely proportional to the transition-matrix prefactor $k_0$, transit-time distributions for different models can be compared by scaling $t_{\text{AB}}$ according to the mean transit time, ${\langle t_{\text{AB}} \rangle}$.
In this way we can see that the predicted distribution of transit times, $p(t_{\text{AB}})$, between the folded and unfolded free-energy minima $x_{\text{A}}$ and $x_{\text{B}}$ (\figref{fig:landscapes}c) is relatively narrow, with a coefficient of variation of $0.39$ and an exponential tail (\figref{fig:tp_times}a).
Alternatively, we can fit the decay constant of the exponential tail, ${\omega^{-1}\!}$, in order to compare with the theoretical distribution for \mbox{1-d} harmonic barrier crossings, ${P_{\text{AB}}^{\text{harm}}(\omega t; \Delta F^\dagger)}$~\cite{chaudhury2010harmonic}, where the shape parameter ${\Delta F^\dagger}$ is the height of the barrier (\figref{fig:tp_times}b).
The simulated distribution of \mbox{1-d} transit times agrees well with this harmonic prediction using the barrier height ${\Delta F^\dagger = 5.37 kT}$ determined directly from the empirical 1-d landscape (\figref{fig:landscapes}a), despite the fact that this landscape is not perfectly harmonic.

We then repeated these calculations using the topological configuration rate matrix defined in \eqref{eq:rates}.
The resulting distribution of transit times (\figref{fig:tp_times}a) also has an exponential tail, but is substantially broader, with a coefficient of variation of $0.91$.
We find that the distribution of transit times obtained from the MD simulations, which has a coefficient of variation of ${1.21 \pm 0.45}$, is considerably closer to the distribution derived from the topological configuration model than the distribution derived from the \mbox{1-d} model.
(The maximum likelihood ratio for the two models given the ten MD transit times is $10^{19}$).
When comparing the transit-time distribution predicted by the topological configuration model with the harmonic prediction ${P_{\text{AB}}^{\text{harm}}(\omega t; \Delta F^\dagger)}$, the best fit is obtained with a shape parameter $\Delta F^\dagger$ that corresponds to a one-dimensional landscape with a ${0.43 kT}$ barrier (\figref{fig:tp_times}b).
Interestingly, this order-of-magnitude difference between the actual barrier height\footnote{Note that the barrier on the minimum free-energy path through the topological configuration network is also greater than ${5 kT}$.} and that returned by a fit to the harmonic theory is reminiscent of the discrepancy found in experimental measurements~\cite{neupane2016direct}.

\begin{figure}
  \includegraphics{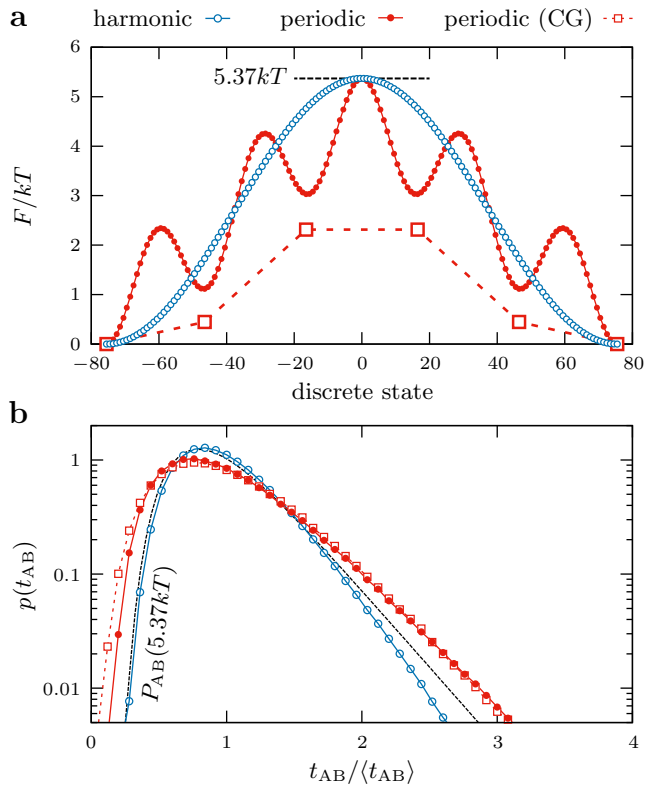}
  \caption{Distributions of transition-path transit times for two one-dimensional toy landscapes.  (a)~The two toy free-energy landscapes, with a single harmonic barrier [blue empty circles; ${F(x) \propto A \cos(2\pi x/151)}$] and five intermediate barriers [red filled circles; ${F(x) \propto A \cos(2\pi x/151) + 2 kT \cos(10\pi x/151)}$], respectively.  The maximum barrier height and number of discrete states were chosen to match the empirical \mbox{1-d} landscape shown in \figref{fig:landscapes}a for each toy landscape.  Coarse-graining (CG) the intermediate-barrier landscape by calculating the mean first passage times between local free-energy minima results in the six-state model shown by red empty squares.  (b)~The transit-time distributions for the two toy landscapes, with all times scaled by the mean transit time, ${\langle t_{\text{AB}} \rangle}$, for each model.  The single-barrier landscape has a transit-time distribution that is narrower than the theoretical prediction, ${P_{\text{AB}}^{\text{harm}}\!}$, for this barrier height (black dashed line), while the intermediate-barrier transit-time distribution and its coarse-grained approximation are significantly broader.}
  \label{fig:toy_tp_times}
\end{figure}

The striking difference between the shapes of these transit-time distributions is primarily a consequence of the intermediate barrier crossings, as opposed to the multidimensionality of the network model.
For example, by simulating transition paths that only traverse the quasi-one-dimensional minimum free-energy path through the configuration network, we obtained a similar distribution of transit times (\figref{fig:tp_times}).
To explore this reasoning further, we constructed two toy \mbox{1-d} landscapes with the same barrier height and number of bins as in the empirical \mbox{1-d} landscape (\figref{fig:toy_tp_times}a).
In the first landscape, the single barrier is, to a good approximation, harmonic, while in the second landscape, there are five intermediate barriers.
We then computed $\bm{T^{\text{1-d}}}$ using \eqref{eq:rates1d} and simulated the transition paths for these toy models via kMC.
As expected, the presence of the intermediate barriers significantly broadens the transit-time distribution, resulting in a coefficient of variation of 0.47 for the intermediate-barrier landscape versus 0.36 for the single-barrier toy landscape (\figref{fig:toy_tp_times}b).
It is also possible to coarse-grain the dynamics over the intermediate-barrier toy landscape by calculating the mean first passage times between the local free-energy minima (\figref{fig:toy_tp_times}a).
Simulating the transition paths for this coarse-grained model results in a good agreement with the full intermediate-barrier toy landscape (\figref{fig:toy_tp_times}b).
Although the difference between the transit-time distributions for these particular toy models is smaller than that shown in \figref{fig:tp_times}, this comparison clearly demonstrates that the presence of intermediate barriers on the transition paths tends to broaden the distribution of transit times.

\subsection{Distribution of finite-time displacements on a one-dimensional order parameter}

\begin{figure}
  \includegraphics{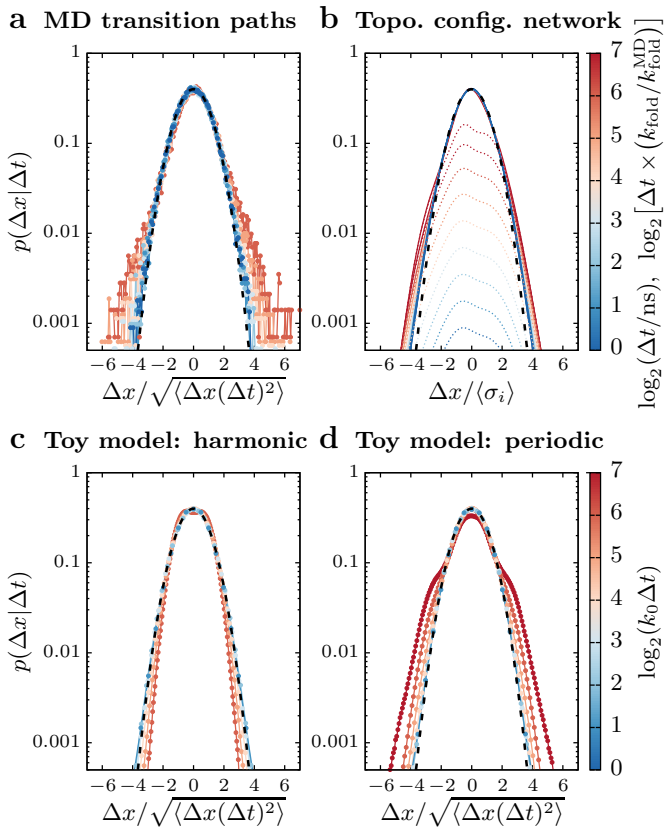}
  \caption{Distributions of finite-time displacements on the \mbox{1-d} native contacts order parameter, averaged over the transition-path ensemble.  (a)~The distribution of displacements $\Delta x$ after a lag time $\Delta t$ observed in all-atom MD transition paths.  The distributions are centered, such that ${\langle \Delta x \rangle = 0}$, and scaled by the root-mean-squared displacement at each lag time.  The frequent small displacements are well described by a Gaussian distribution with a unit standard deviation (black dashed line); however, larger displacements are much more frequent than predicted by the tails of this Gaussian distribution.  Colors correspond to the lag time, in units of nanoseconds, as shown by the scale bar on the right.  (b)~The predicted distribution of displacements corresponding to transitions \textit{between} configurations in the topological configuration model (dotted lines) is broader than a Gaussian distribution fit to the transition-path-ensemble-averaged fluctuations \textit{within} individual configurations, leading to similar fat-tailed behavior.  The lag times are scaled by the slowest timescale of the MD simulations, ${(k_{\text{fold}}^{\text{MD}})^{-1}\!}$, for comparison with panel a.  (c--d)~Distributions calculated from kinetic Monte Carlo simulations of transition paths on the two toy landscapes shown in \figref{fig:toy_tp_times}a.  Only the intermediate-barrier landscape (panel d) reproduces the fat-tailed behavior observed in the MD simulations.}
  \label{fig:step_sizes}
\end{figure}

As a second statistical test, we examined distributions of finite-time displacements on a one-dimensional order parameter.
Using the number of native contacts as the order parameter, we measured displacements, ${\Delta x}$, given a lag time ${\Delta t}$ on all transition-path trajectories in the atomistic MD simulations.
We considered lag times ranging from ${1\,\text{ns}}$, which is longer than the typical time required for the formation of a single native contact, to ${\sim 100\,\text{ns}}$, which is much shorter than the mean transit time, ${2.43\,\mu\text{s}}$, observed in the MD simulations.
After verifying that Ubiquitin transition paths exhibit subdiffusive motion over this range of lag times~\cite{krivov2010protein}, meaning that ${\langle [\Delta x(\Delta t)]^2 \rangle \propto (\Delta t)^p}$ with ${p < 1}$, we sought to determine whether, for a given lag time, the distribution of frequent, small displacements is predictive of larger jumps.
By averaging over all MD transition-path trajectories and removing the net directional motion, ${(x_{\text{B}} - x_{\text{A}}) / t_{\text{AB}}}$, we find that the vast majority of displacements, for which ${|\Delta x| \le 2 \sqrt{\langle \Delta x(\Delta t)^2 \rangle}}$, are well described by Gaussian distributions over the entire range of lag times.
However, larger displacements are much more frequent than predicted by the tails of the Gaussian distributions, regardless of the lag time.
This `fat-tailed' behavior is shown in \figref{fig:step_sizes}a, where the distribution for each lag time is scaled according to its root-mean-squared displacement and compared to a unit Gaussian distribution indicated by the black dashed line.

This unusual feature is naturally predicted by the timescale separation in the topological configuration model, since the distribution of one-dimensional displacements is narrower for fluctuations within a configuration than for transitions from one configuration to another.
To illustrate this idea, we compare the expected displacement associated with a step between configurations, ${\langle (\sigma_i^2 + \sigma_j^2)^{1/2} \rangle}$, with the average size of a fluctuation within a configuration, ${\langle \sigma_i^2 \rangle}$, where the averages are taken over all configurations and weighted by ${m^{\text{AB}}_i\!}$, the probability of finding a transition-path trajectory in any configuration $i$ (\figref{fig:step_sizes}b).
The contribution from transitions between configurations, shown by colored dotted lines in \figref{fig:step_sizes}b, increases with the lag time, since the probability of moving from configuration $i$ to $j$ within a finite time $\Delta t$ is given by the matrix exponential ${[\exp(\Delta t \bm{T})]_{ij}}$.
As a result of these larger displacements, the tails of the distribution are always outside of the unit Gaussian distribution, suggesting that the fat-tailed behavior observed in the MD simulation-derived distributions is also indicative of relatively rare intermediate barrier crossings.

To test this hypothesis, we analyzed the distributions of finite-time displacements obtained from simulated transition paths over the two toy landscapes shown in \figref{fig:toy_tp_times}a.
By scaling the distributions according to their root-mean-squared displacements and comparing with a unit Gaussian distribution (\figref{fig:step_sizes}c,d), we find that only the landscape with intermediate barriers results in a qualitatively similar fat-tailed distribution.
The single-barrier toy landscape, by contrast, results in large displacements being less frequent than predicted by a Gaussian distribution.
We therefore conclude that the relative enrichment of large displacements within lag times of a few tens of nanoseconds is a likely consequence of intermediate barrier crossings in the MD-simulated transition paths.

\subsection{Likelihood comparison between predicted and simulated transition-path trajectories}

Finally, we tested whether the transition-path trajectories observed in the MD simulations, when projected onto a one-dimensional coordinate, are representative of the transition-path ensembles predicted by the topological configuration model.
To do so, we treated this stochastic process as a hidden Markov model with a discrete state space of topological configurations.
In this model, transition paths traverse the discrete state space in accordance with the rate matrix $\bm{T}$, but we assume that we can only observe the instantaneous projection of each state $s$ onto the \mbox{1-d} coordinate $x$.
With the exception of the transition-matrix prefactor $k_0$, both the transition probabilities between states and the topological configuration-dependent probabilities of observing a given number of native contacts, $p(x|s)$, are completely determined by quantities calculated from the equilibrium MD simulation data.
We first removed all configurations in the topological configuration network that are not likely to be visited on transition paths (less than 1\% of predicted folding flux) to guard against overfitting.
We then used the standard Viterbi algorithm~\cite{forney1973viterbi,hmmlearn} to determine the unique sequence of configurations, ${\{s_l\}}$, that maximizes the log likelihood of the observed time series ${\{x_l\}}$ for each transition-path trajectory,
\begin{equation}
  \langle \log L \rangle = n^{-1} \sum_{l=1}^n \log\!\left[ p(x_l | s_l) ( e^{\Delta t \bm{T}} )_{s_{l-1},s_l} \right],
\end{equation}
where the index $l$ runs over all consecutive snapshots on each transition path, and we have normalized the log likelihood to remove the trivial dependence on the total trajectory length~$n$.

\begin{figure}
  \includegraphics{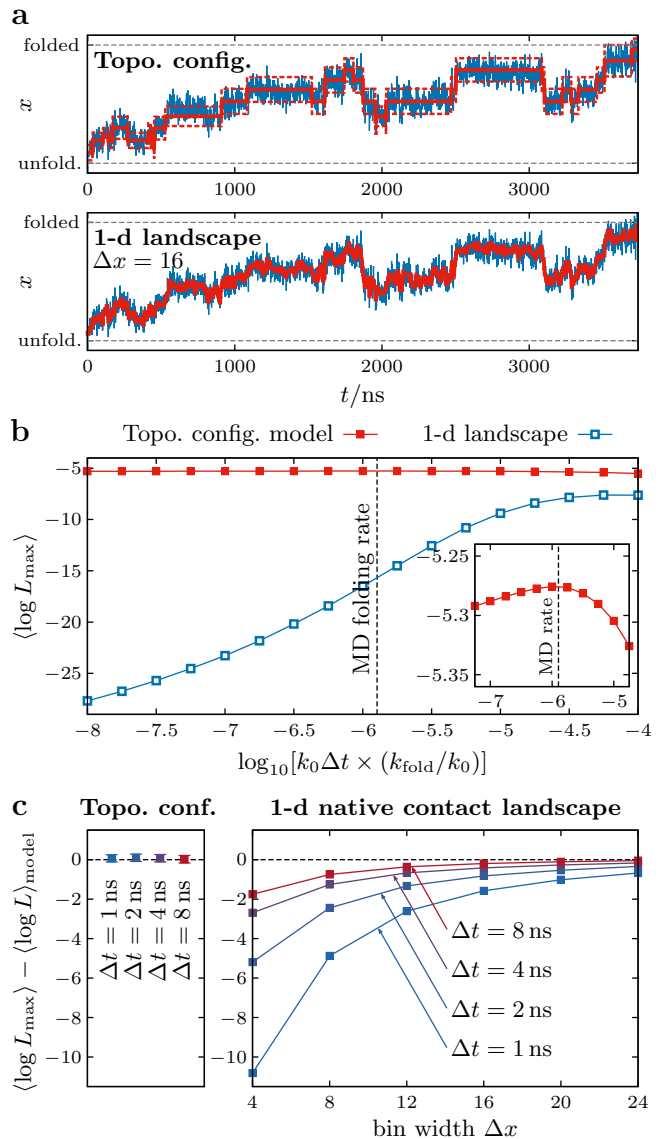}
  \caption{Comparison of transition-path trajectories from atomistic MD simulations and trajectories generated by theoretical models.  (a)~The maximum-likelihood transition paths (red solid lines) through the discrete states of the topological configuration and \mbox{1-d} landscape models given a representative MD transition-path trajectory, projected onto the number of native contacts, $x$ (blue lines).  For the topological configuration model, the expected fluctuations (i.e., the standard deviations of the Gaussian approximations in \figref{fig:landscapes}b) within the most probable states are shown by red dashed lines.  (b)~The dependence of the per-frame log likelihood of the most probable sequence of states on the transition-matrix prefactor, $k_0$.  This rate is scaled by the folding rate predicted by the model, ${k_{\text{fold}}}$, for comparison with the estimated MD folding rate, ${k_{\text{fold}}^{\text{MD}}}$ (black dashed line).  The maximum of ${\langle \log L_{\text{max}} \rangle}$ for the topological configuration model coincides with the MD folding rate (inset).  (c)~Comparison between the log likelihood of the most probable sequence of states and the expected log likelihood for each model, ${\langle \log L \rangle_{\text{model}}}$ (see text).  When fitting to the \mbox{1-d} model, the agreement between these quantities depends strongly on the spatial coarse-graining, $\Delta x$, of the landscape and temporal averaging of the MD transition-path trajectories over time windows of width $\Delta t$.  Error bars indicate the standard error of the mean.}
  \label{fig:trajectories}
\end{figure}

A representative maximum likelihood fit, using the fixed transition rates and emission probabilities defined in \eqref{eq:rates} and \figref{fig:landscapes}b, respectively, is shown in \figref{fig:trajectories}a, where the apparent separation of timescales between high-frequency oscillations and slower, step-like behavior can be easily discerned by eye.
Furthermore, the log likelihood of most probable path, ${\langle \log L_{\text{max}} \rangle}$, is only weakly dependent on the transition-matrix prefactor, $k_0$. (\figref{fig:trajectories}b).
Analyzing one frame per nanosecond, the maximum of ${\langle \log L_{\text{max}} \rangle}$ is found at a value of $k_0$ that results in a predicted folding rate, $k_{\text{fold}}$, of approximately ${1\times10^{-6}}$ frames.
Importantly, this folding rate agrees well with the empirical folding rate that we calculated directly from the mean waiting times between folding and unfolding transitions in the full MD trajectories, ${k_{\text{fold}}^{\text{MD}} \simeq 1.3\,\text{ms}^{-1}}$ (\figref{fig:trajectories}b,\textit{inset}).

We then performed an analogous hidden Markov analysis for the \mbox{1-d} landscape model with a constant diffusion coefficient.
In this case, the discrete states ${\{\bar{x}\}}$ are bins of width ${\Delta x}$, the rate matrix is given by \eqref{eq:rates1d}, and the Gaussian emissions ${p(x|s=\bar{x})}$ are assumed to have a standard deviation of ${\Delta x}$.
For example, a representative maximum likelihood fit, assuming a bin width of ${\Delta x = 16}$ native contacts, is shown in \figref{fig:trajectories}a.
Unlike the topological configuration model, we find that the log likelihood of the most probable path in the \mbox{1-d} model is strongly dependent on the transition-matrix prefactor, $k_0$.
Furthermore, the maximum with respect to $k_0$ corresponds to a folding rate that is orders of magnitude greater than that determined from MD simulations.
This means that, in order to generate a transition path with the observed high frequency fluctuations on the empirical \mbox{1-d} landscape, $k_0$ has to be tuned to a point where the predicted rates of folding and unfolding events are unrealistically fast.
The topological configuration model does not suffer from this contradiction, since the separation of timescales between the fast motions within a topological configuration and slower transitions between configurations is an intrinsic feature of the model.

To ensure a fair comparison between these models, we calculated the expected values of the log likelihood for transition paths generated directly by both models,
\begin{equation}
  \langle \log L \rangle_{\text{model}} = \sum_{s} m^{\text{AB}}_s \log\!\left[ \langle p(x | s) \rangle \langle ( e^{\Delta t \bm{T}} )_{s'\!,s} \rangle \right],
\end{equation}
where the expectation values for the emission and transition probabilities are ${\langle p(x | s) \rangle \equiv \int_{-\infty}^{\infty} p(x|s)^2 dx}$ and ${\langle [\exp( \Delta t \bm{T} )]_{s'\!,s} \rangle \equiv \sum_{s'} \{[\exp( \Delta t \bm{T} )]_{s'\!,s}\}^2}$, respectively, in each state $s$.
Fixing $k_0$ to match the MD folding rate, \figref{fig:trajectories}c shows that the log likelihood of the most probable path determined by fitting the MD data is consistent with the expected log likelihood for the ensemble of transition paths generated by the topological configuration model.
This result is independent of temporal coarse-graining, i.e., down-sampling the trajectory ${x(t)}$ by averaging over a moving window of width $\Delta t$.
By contrast, temporal coarse-graining has a significant effect on the difference between the best-fit and expected log likelihoods for the \mbox{1-d} landscape model when $k_0$ is fixed according to the MD folding rate, since increasing $\Delta t$ preferentially removes high frequency fluctuations.
This difference is also sensitive to the landscape bin width $\Delta x$, since increasing $\Delta x$ reduces the number of distinct states and consequently slows the rate of transitions between adjacent states.
As a result, increasing the bin width results in an effective separation of timescales on the \mbox{1-d} landscape, albeit without a first-principles justification.
Only by introducing a separation of timescales through a \textit{post hoc} combination of temporal coarse-graining of the trajectories and spatial coarse-graining of the \mbox{1-d} landscape is it possible for the 1-d model to generate transition paths that are consistent with those observed in the MD simulations (\figref{fig:trajectories}c).

In conclusion, this hidden Markov analysis highlights the importance of a separation of timescales for reproducing the transition-path trajectories observed in atomistic MD simulations.
In particular, the co-occurrence of fast fluctuations in the number of native contacts and infrequent folding events is naturally captured by the topological configuration model, as seen by the agreement between the log likelihoods of the fitted and predicted transition-path trajectory ensembles.
One-dimensional landscape models that lack an intermediate timescale, by contrast, require that the trajectories observed in MD simulations be smoothed substantially in order to conform to the predicted transition-path dynamics.

\section{Discussion}

We have shown that a theoretical model of protein folding, which emphasizes an intermediate timescale associated with transitions between distinct configurations of partial native structure, accurately predicts multiple statistical properties of the stochastic dynamics of folding transition paths.
By using equilibrium simulation data to construct an approximate rate matrix for transitions between topological configurations, we demonstrated that the transition-path ensembles generated by this model have broad transit-time distributions that are consistent with both all-atom simulations and experimental observations.
We then showed that the non-Gaussian distributions of finite-time displacements that are predicted by this model qualitatively match all-atom simulation results.
Lastly, we demonstrated that this model can reconcile rapid local fluctuations on a one-dimensional order parameter with a slow overall rate of folding, two seemingly contradictory features that are simultaneously observed in simulated transition-path trajectories.
Most importantly, all predictions of the topological configuration model were made without the use of any dynamical information.

The intermediate timescale in the topological configuration model is predicted to arise due to local free-energy barriers that separate transient states with distinct sets of native-like loops~\cite{jacobs2016structure}.
In this work, we assumed that the rates of transitions between states could be approximated using a simple formula that satisfies detailed balance.
However, a more accurate approach would involve the calculation of the rates between adjacent topological configurations in the all-atom model.
Such an approach might benefit from recent advances in Markov state modeling~\cite{bowman2014introduction,chodera2014markov}, although, in this application, the definitions of the states would be assumed \textit{a priori} on the basis of the native structure.
Nevertheless, it is remarkable that we are able to obtain qualitatively accurate results for a variety of statistical tests using a highly simplified Markov model and the estimated free energies of the predicted topological configurations.
This success indicates that the statistical analyses that we considered depend to a greater extent on the presence of intermediate barriers than on their precise heights, provided that these barriers are comparable to the thermal energy (${\gtrsim \kB T}$) and are thus kinetically relevant.
Generalizing beyond Ubiquitin, we anticipate that accounting for this intermediate timescale is likely to be especially important in the context of large proteins, which tend to have complicated native topologies.
Transition-path analyses of small, ultrafast-folding proteins~\cite{lindorff2011fast}, by contrast, are less likely to benefit from the coarse-graining strategy described here, since these proteins typically contain only one or two native-state loops whose formation dominates the overall folding rate.
The lower probability of encountering substantial free-energy barriers on folding transition paths, which are needed to assume approximately Markovian transitions between intermediate configurations, suggests that models of diffusion over a single harmonic barrier may be more appropriate in these particular cases.

To compare our approach with commonly used one-dimensional models, we assumed a projection onto the number of native contacts and a constant diffusion coefficient.
The number of native contacts has been shown to be a good reaction coordinate in the sense that, for many small proteins, it can be used to locate the transition state from transition-path trajectories with high probability~\cite{best2013native}.
The results presented here are not inconsistent with this notion of a good reaction coordinate when a single pathway through the network of topological configurations dominates, as shown by the similarity between the committors predicted by the two models.
Furthermore, if one were to account for coordinate-dependent diffusion, it is likely that these two approaches would lead to similar predictions for the transition-path dynamics, since existing methods for fitting coordinate-dependent diffusion coefficients often reveal the existence of intermediate barriers that were hidden by the projection onto the original reaction coordinate~\cite{best2010coordinate}.
However, to carry out such an analysis, dynamical information is always required in some form~\cite{best2011diffusion}, meaning that the underlying landscape is not, by itself, truly predictive.

The topological configuration model that we have examined here differs in a number of important ways from alternative models of folding intermediates that have been proposed previously.
Unlike the early hierarchical model of Ptitsyn~\cite{ptitsyn1973stages} and the more recent `foldon' hypothesis~\cite{maity2005protein}, the assembly of native-like intermediates need not lead to a more negative free energy at every step.
At the same time, we have not assumed that the free energy decreases only upon incorporation of the final native-like substructure, as proposed in a recent `volcano' model of folding~\cite{rollins2014general}.
By contrast, the highest point on the minimum free-energy path through the network of topological configurations is determined by the free energies of the various configurations and the barriers between them, which depend, in turn, on the temperature and solvent conditions~\cite{jacobs2016structure}.
The topological configuration model also suggests a natural definition of a folding pathway~\cite{adhikari2012novo} at the level of topological configurations while allowing for alternative, yet less probable, pathways.

Finally, this theoretical analysis has a number of implications for experimental investigations of protein-folding transition paths.
The transit-time and finite-time displacement distributions that we have discussed can now be measured directly in single-molecule experiments.
However, to distinguish between alternative theoretical models, it would be most useful to analyze high-resolution experimental transition-path measurements using hidden Markov models in order to detect and characterize transient folding intermediates.
Using established non-parametric methods~\cite{sgouralis2018single,sgouralis2017icon}, it should be possible to assess both the number of distinct transient states and any separation of timescales objectively.
Furthermore, by combining such analyses with structure-based models, like the type discussed here, it should be possible to extract more detailed information regarding the underlying free-energy landscape from these measurements.
In this way, continued advances in single-molecule measurements can be used to improve predictive landscape-based models of protein-folding transition paths, in particular for large proteins with complex native-state topologies.

\section{Conclusion}

We have shown that the introduction of an intermediate timescale, which is faster than native-contact formation but slower than the typical time for folding an entire protein, can qualitatively alter the statistics of protein-folding transition paths.
We proposed that this intermediate timescale is associated with the assembly of native-like loops, and we used this principle to build a coarse-grained free-energy landscape for Ubiquitin from equilibrium atomistic simulation data.
Without relying on any dynamical information from simulations, we showed that this model predicts distributions of transit times and finite-time displacements that are consistent with simulated transition paths, but differ qualitatively from the predictions of a one-dimensional model of diffusion on an empirical free-energy landscape.
We also used a hidden Markov analysis to demonstrate that this model generates transition paths that agree with both the dynamics and kinetics inferred from reversible folding simulations.
Our results suggest that the analysis of single-molecule transition-path trajectories may be improved by accounting for intermediate free-energy barriers, which are a fundamental aspect of the complexity of folding large biomolecules.

\begin{acknowledgments}
The authors would like to thank D.E.~Shaw Research for providing the all-atom Molecular Dynamics simulation data.
In addition, the authors are grateful for many insightful discussions with Michael Manhart.
This work was supported by NIH grants F32GM116231 and GM068670.
All analysis and simulation code is available from the authors upon request.
\end{acknowledgments}


\begin{thebibliography}{70}%
\makeatletter
\providecommand \@ifxundefined [1]{%
 \@ifx{#1\undefined}
}%
\providecommand \@ifnum [1]{%
 \ifnum #1\expandafter \@firstoftwo
 \else \expandafter \@secondoftwo
 \fi
}%
\providecommand \@ifx [1]{%
 \ifx #1\expandafter \@firstoftwo
 \else \expandafter \@secondoftwo
 \fi
}%
\providecommand \natexlab [1]{#1}%
\providecommand \enquote  [1]{``#1''}%
\providecommand \bibnamefont  [1]{#1}%
\providecommand \bibfnamefont [1]{#1}%
\providecommand \citenamefont [1]{#1}%
\providecommand \href@noop [0]{\@secondoftwo}%
\providecommand \href [0]{\begingroup \@sanitize@url \@href}%
\providecommand \@href[1]{\@@startlink{#1}\@@href}%
\providecommand \@@href[1]{\endgroup#1\@@endlink}%
\providecommand \@sanitize@url [0]{\catcode `\\12\catcode `\$12\catcode
  `\&12\catcode `\#12\catcode `\^12\catcode `\_12\catcode `\%12\relax}%
\providecommand \@@startlink[1]{}%
\providecommand \@@endlink[0]{}%
\providecommand \url  [0]{\begingroup\@sanitize@url \@url }%
\providecommand \@url [1]{\endgroup\@href {#1}{\urlprefix }}%
\providecommand \urlprefix  [0]{URL }%
\providecommand \Eprint [0]{\href }%
\providecommand \doibase [0]{http://dx.doi.org/}%
\providecommand \selectlanguage [0]{\@gobble}%
\providecommand \bibinfo  [0]{\@secondoftwo}%
\providecommand \bibfield  [0]{\@secondoftwo}%
\providecommand \translation [1]{[#1]}%
\providecommand \BibitemOpen [0]{}%
\providecommand \bibitemStop [0]{}%
\providecommand \bibitemNoStop [0]{.\EOS\space}%
\providecommand \EOS [0]{\spacefactor3000\relax}%
\providecommand \BibitemShut  [1]{\csname bibitem#1\endcsname}%
\let\auto@bib@innerbib\@empty
\bibitem [{\citenamefont {Frenkel}\ and\ \citenamefont
  {Smit}(2001)}]{frenkel2001understanding}%
  \BibitemOpen
  \bibfield  {author} {\bibinfo {author} {\bibfnamefont {D.}~\bibnamefont
  {Frenkel}}\ and\ \bibinfo {author} {\bibfnamefont {B.}~\bibnamefont {Smit}},\
  }\href@noop {} {\emph {\bibinfo {title} {Understanding molecular simulation:
  {F}rom algorithms to applications}}}\ (\bibinfo  {publisher} {Academic
  Press},\ \bibinfo {year} {2001})\BibitemShut {NoStop}%
\bibitem [{\citenamefont {Gibbs}(1878)}]{gibbs1878equilibrium}%
  \BibitemOpen
  \bibfield  {author} {\bibinfo {author} {\bibfnamefont {J.~W.}\ \bibnamefont
  {Gibbs}},\ }\href@noop {} {\bibfield  {journal} {\bibinfo  {journal} {Am. J.
  Sci.}\ }\textbf {\bibinfo {volume} {96}},\ \bibinfo {pages} {441} (\bibinfo
  {year} {1878})}\BibitemShut {NoStop}%
\bibitem [{\citenamefont {ten Wolde}\ \emph {et~al.}(1996)\citenamefont {ten
  Wolde}, \citenamefont {Ruiz-Montero},\ and\ \citenamefont
  {Frenkel}}]{rein1996numerical}%
  \BibitemOpen
  \bibfield  {author} {\bibinfo {author} {\bibfnamefont {P.~R.}\ \bibnamefont
  {ten Wolde}}, \bibinfo {author} {\bibfnamefont {M.~J.}\ \bibnamefont
  {Ruiz-Montero}}, \ and\ \bibinfo {author} {\bibfnamefont {D.}~\bibnamefont
  {Frenkel}},\ }\href@noop {} {\bibfield  {journal} {\bibinfo  {journal} {J.
  Chem. Phys.}\ }\textbf {\bibinfo {volume} {104}},\ \bibinfo {pages} {9932}
  (\bibinfo {year} {1996})}\BibitemShut {NoStop}%
\bibitem [{\citenamefont {De~Yoreo}\ and\ \citenamefont
  {Vekilov}(2003)}]{de2003principles}%
  \BibitemOpen
  \bibfield  {author} {\bibinfo {author} {\bibfnamefont {J.~J.}\ \bibnamefont
  {De~Yoreo}}\ and\ \bibinfo {author} {\bibfnamefont {P.~G.}\ \bibnamefont
  {Vekilov}},\ }\href@noop {} {\bibfield  {journal} {\bibinfo  {journal} {Rev.
  Mineral. Geochem.}\ }\textbf {\bibinfo {volume} {54}},\ \bibinfo {pages} {57}
  (\bibinfo {year} {2003})}\BibitemShut {NoStop}%
\bibitem [{\citenamefont {ten Wolde}\ and\ \citenamefont
  {Frenkel}(1997)}]{ten1997enhancement}%
  \BibitemOpen
  \bibfield  {author} {\bibinfo {author} {\bibfnamefont {P.~R.}\ \bibnamefont
  {ten Wolde}}\ and\ \bibinfo {author} {\bibfnamefont {D.}~\bibnamefont
  {Frenkel}},\ }\href@noop {} {\bibfield  {journal} {\bibinfo  {journal}
  {Science}\ }\textbf {\bibinfo {volume} {277}},\ \bibinfo {pages} {1975}
  (\bibinfo {year} {1997})}\BibitemShut {NoStop}%
\bibitem [{\citenamefont {Jacobs}\ \emph {et~al.}(2015)\citenamefont {Jacobs},
  \citenamefont {Reinhardt},\ and\ \citenamefont
  {Frenkel}}]{jacobs2015rational}%
  \BibitemOpen
  \bibfield  {author} {\bibinfo {author} {\bibfnamefont {W.~M.}\ \bibnamefont
  {Jacobs}}, \bibinfo {author} {\bibfnamefont {A.}~\bibnamefont {Reinhardt}}, \
  and\ \bibinfo {author} {\bibfnamefont {D.}~\bibnamefont {Frenkel}},\
  }\href@noop {} {\bibfield  {journal} {\bibinfo  {journal} {Proc. Natl. Acad.
  Sci. U.S.A.}\ }\textbf {\bibinfo {volume} {112}},\ \bibinfo {pages} {6313}
  (\bibinfo {year} {2015})}\BibitemShut {NoStop}%
\bibitem [{\citenamefont {Frenkel}(1999)}]{frenkel1999entropy}%
  \BibitemOpen
  \bibfield  {author} {\bibinfo {author} {\bibfnamefont {D.}~\bibnamefont
  {Frenkel}},\ }\href@noop {} {\bibfield  {journal} {\bibinfo  {journal}
  {Physica A: Stat. Mech.}\ }\textbf {\bibinfo {volume} {263}},\ \bibinfo
  {pages} {26} (\bibinfo {year} {1999})}\BibitemShut {NoStop}%
\bibitem [{\citenamefont {Poon}\ \emph {et~al.}(2000)\citenamefont {Poon},
  \citenamefont {Renth},\ and\ \citenamefont {Evans}}]{poon2000kinetics}%
  \BibitemOpen
  \bibfield  {author} {\bibinfo {author} {\bibfnamefont {W.}~\bibnamefont
  {Poon}}, \bibinfo {author} {\bibfnamefont {F.}~\bibnamefont {Renth}}, \ and\
  \bibinfo {author} {\bibfnamefont {R.}~\bibnamefont {Evans}},\ }\href@noop {}
  {\bibfield  {journal} {\bibinfo  {journal} {J. Phys.: Condens. Matter}\
  }\textbf {\bibinfo {volume} {12}},\ \bibinfo {pages} {A269} (\bibinfo {year}
  {2000})}\BibitemShut {NoStop}%
\bibitem [{\citenamefont {Abkevich}\ \emph
  {et~al.}(1994{\natexlab{a}})\citenamefont {Abkevich}, \citenamefont {Gutin},\
  and\ \citenamefont {Shakhnovich}}]{abkevich1994free}%
  \BibitemOpen
  \bibfield  {author} {\bibinfo {author} {\bibfnamefont {V.}~\bibnamefont
  {Abkevich}}, \bibinfo {author} {\bibfnamefont {A.}~\bibnamefont {Gutin}}, \
  and\ \bibinfo {author} {\bibfnamefont {E.}~\bibnamefont {Shakhnovich}},\
  }\href@noop {} {\bibfield  {journal} {\bibinfo  {journal} {J. Chem. Phys.}\
  }\textbf {\bibinfo {volume} {101}},\ \bibinfo {pages} {6052} (\bibinfo {year}
  {1994}{\natexlab{a}})}\BibitemShut {NoStop}%
\bibitem [{\citenamefont {Shakhnovich}(2006)}]{shakhnovich2006protein}%
  \BibitemOpen
  \bibfield  {author} {\bibinfo {author} {\bibfnamefont {E.~I.}\ \bibnamefont
  {Shakhnovich}},\ }\href@noop {} {\bibfield  {journal} {\bibinfo  {journal}
  {Chem. Rev.}\ }\textbf {\bibinfo {volume} {106}},\ \bibinfo {pages} {1559}
  (\bibinfo {year} {2006})}\BibitemShut {NoStop}%
\bibitem [{\citenamefont {Gfeller}\ \emph {et~al.}(2007)\citenamefont
  {Gfeller}, \citenamefont {De~Los~Rios}, \citenamefont {Caflisch},\ and\
  \citenamefont {Rao}}]{gfeller2007complex}%
  \BibitemOpen
  \bibfield  {author} {\bibinfo {author} {\bibfnamefont {D.}~\bibnamefont
  {Gfeller}}, \bibinfo {author} {\bibfnamefont {P.}~\bibnamefont
  {De~Los~Rios}}, \bibinfo {author} {\bibfnamefont {A.}~\bibnamefont
  {Caflisch}}, \ and\ \bibinfo {author} {\bibfnamefont {F.}~\bibnamefont
  {Rao}},\ }\href@noop {} {\bibfield  {journal} {\bibinfo  {journal} {Proc.
  Natl. Acad. Sci. U.S.A.}\ }\textbf {\bibinfo {volume} {104}},\ \bibinfo
  {pages} {1817} (\bibinfo {year} {2007})}\BibitemShut {NoStop}%
\bibitem [{\citenamefont {Thirumalai}\ \emph {et~al.}(2010)\citenamefont
  {Thirumalai}, \citenamefont {O'Brien}, \citenamefont {Morrison},\ and\
  \citenamefont {Hyeon}}]{thirumalai2010theoretical}%
  \BibitemOpen
  \bibfield  {author} {\bibinfo {author} {\bibfnamefont {D.}~\bibnamefont
  {Thirumalai}}, \bibinfo {author} {\bibfnamefont {E.~P.}\ \bibnamefont
  {O'Brien}}, \bibinfo {author} {\bibfnamefont {G.}~\bibnamefont {Morrison}}, \
  and\ \bibinfo {author} {\bibfnamefont {C.}~\bibnamefont {Hyeon}},\
  }\href@noop {} {\bibfield  {journal} {\bibinfo  {journal} {Ann. Rev.
  Biophys.}\ }\textbf {\bibinfo {volume} {39}},\ \bibinfo {pages} {159}
  (\bibinfo {year} {2010})}\BibitemShut {NoStop}%
\bibitem [{\citenamefont {Abkevich}\ \emph
  {et~al.}(1994{\natexlab{b}})\citenamefont {Abkevich}, \citenamefont {Gutin},\
  and\ \citenamefont {Shakhnovich}}]{abkevich1994specific}%
  \BibitemOpen
  \bibfield  {author} {\bibinfo {author} {\bibfnamefont {V.~I.}\ \bibnamefont
  {Abkevich}}, \bibinfo {author} {\bibfnamefont {A.~M.}\ \bibnamefont {Gutin}},
  \ and\ \bibinfo {author} {\bibfnamefont {E.~I.}\ \bibnamefont
  {Shakhnovich}},\ }\href@noop {} {\bibfield  {journal} {\bibinfo  {journal}
  {Biochemistry}\ }\textbf {\bibinfo {volume} {33}},\ \bibinfo {pages} {10026}
  (\bibinfo {year} {1994}{\natexlab{b}})}\BibitemShut {NoStop}%
\bibitem [{\citenamefont {Shakhnovich}(1997)}]{shakhnovich1997theoretical}%
  \BibitemOpen
  \bibfield  {author} {\bibinfo {author} {\bibfnamefont {E.~I.}\ \bibnamefont
  {Shakhnovich}},\ }\href@noop {} {\bibfield  {journal} {\bibinfo  {journal}
  {Curr. Opin. Struct. Biol.}\ }\textbf {\bibinfo {volume} {7}},\ \bibinfo
  {pages} {29} (\bibinfo {year} {1997})}\BibitemShut {NoStop}%
\bibitem [{\citenamefont {Chung}\ \emph {et~al.}(2012)\citenamefont {Chung},
  \citenamefont {McHale}, \citenamefont {Louis},\ and\ \citenamefont
  {Eaton}}]{chung2012single}%
  \BibitemOpen
  \bibfield  {author} {\bibinfo {author} {\bibfnamefont {H.~S.}\ \bibnamefont
  {Chung}}, \bibinfo {author} {\bibfnamefont {K.}~\bibnamefont {McHale}},
  \bibinfo {author} {\bibfnamefont {J.~M.}\ \bibnamefont {Louis}}, \ and\
  \bibinfo {author} {\bibfnamefont {W.~A.}\ \bibnamefont {Eaton}},\ }\href@noop
  {} {\bibfield  {journal} {\bibinfo  {journal} {Science}\ }\textbf {\bibinfo
  {volume} {335}},\ \bibinfo {pages} {981} (\bibinfo {year}
  {2012})}\BibitemShut {NoStop}%
\bibitem [{\citenamefont {Chung}\ and\ \citenamefont
  {Eaton}(2013)}]{chung2013single}%
  \BibitemOpen
  \bibfield  {author} {\bibinfo {author} {\bibfnamefont {H.~S.}\ \bibnamefont
  {Chung}}\ and\ \bibinfo {author} {\bibfnamefont {W.~A.}\ \bibnamefont
  {Eaton}},\ }\href@noop {} {\bibfield  {journal} {\bibinfo  {journal}
  {Nature}\ }\textbf {\bibinfo {volume} {502}},\ \bibinfo {pages} {685}
  (\bibinfo {year} {2013})}\BibitemShut {NoStop}%
\bibitem [{\citenamefont {Chung}\ \emph {et~al.}(2013)\citenamefont {Chung},
  \citenamefont {Cellmer}, \citenamefont {Louis},\ and\ \citenamefont
  {Eaton}}]{chung2013measuring}%
  \BibitemOpen
  \bibfield  {author} {\bibinfo {author} {\bibfnamefont {H.~S.}\ \bibnamefont
  {Chung}}, \bibinfo {author} {\bibfnamefont {T.}~\bibnamefont {Cellmer}},
  \bibinfo {author} {\bibfnamefont {J.~M.}\ \bibnamefont {Louis}}, \ and\
  \bibinfo {author} {\bibfnamefont {W.~A.}\ \bibnamefont {Eaton}},\ }\href@noop
  {} {\bibfield  {journal} {\bibinfo  {journal} {Chem. Phys.}\ }\textbf
  {\bibinfo {volume} {422}},\ \bibinfo {pages} {229} (\bibinfo {year}
  {2013})}\BibitemShut {NoStop}%
\bibitem [{\citenamefont {Truex}\ \emph {et~al.}(2015)\citenamefont {Truex},
  \citenamefont {Chung}, \citenamefont {Louis},\ and\ \citenamefont
  {Eaton}}]{truex2015testing}%
  \BibitemOpen
  \bibfield  {author} {\bibinfo {author} {\bibfnamefont {K.}~\bibnamefont
  {Truex}}, \bibinfo {author} {\bibfnamefont {H.~S.}\ \bibnamefont {Chung}},
  \bibinfo {author} {\bibfnamefont {J.~M.}\ \bibnamefont {Louis}}, \ and\
  \bibinfo {author} {\bibfnamefont {W.~A.}\ \bibnamefont {Eaton}},\ }\href@noop
  {} {\bibfield  {journal} {\bibinfo  {journal} {Phys. Rev. Lett.}\ }\textbf
  {\bibinfo {volume} {115}},\ \bibinfo {pages} {018101} (\bibinfo {year}
  {2015})}\BibitemShut {NoStop}%
\bibitem [{\citenamefont {\v{S}ali}\ \emph {et~al.}(1994)\citenamefont
  {\v{S}ali}, \citenamefont {Shakhnovich},\ and\ \citenamefont
  {Karplus}}]{sali1994how}%
  \BibitemOpen
  \bibfield  {author} {\bibinfo {author} {\bibfnamefont {A.}~\bibnamefont
  {\v{S}ali}}, \bibinfo {author} {\bibfnamefont {E.~I.}\ \bibnamefont
  {Shakhnovich}}, \ and\ \bibinfo {author} {\bibfnamefont {M.}~\bibnamefont
  {Karplus}},\ }\href@noop {} {\bibfield  {journal} {\bibinfo  {journal}
  {Nature}\ }\textbf {\bibinfo {volume} {369}},\ \bibinfo {pages} {248}
  (\bibinfo {year} {1994})}\BibitemShut {NoStop}%
\bibitem [{\citenamefont {Bryngelson}\ \emph {et~al.}(1995)\citenamefont
  {Bryngelson}, \citenamefont {Onuchic}, \citenamefont {Socci},\ and\
  \citenamefont {Wolynes}}]{bryngelson1995funnels}%
  \BibitemOpen
  \bibfield  {author} {\bibinfo {author} {\bibfnamefont {J.~D.}\ \bibnamefont
  {Bryngelson}}, \bibinfo {author} {\bibfnamefont {J.~N.}\ \bibnamefont
  {Onuchic}}, \bibinfo {author} {\bibfnamefont {N.~D.}\ \bibnamefont {Socci}},
  \ and\ \bibinfo {author} {\bibfnamefont {P.~G.}\ \bibnamefont {Wolynes}},\
  }\href@noop {} {\bibfield  {journal} {\bibinfo  {journal} {Proteins: Struct.,
  Func., and Bioinf.}\ }\textbf {\bibinfo {volume} {21}},\ \bibinfo {pages}
  {167} (\bibinfo {year} {1995})}\BibitemShut {NoStop}%
\bibitem [{\citenamefont {Onuchic}\ and\ \citenamefont
  {Wolynes}(2004)}]{onuchic2004theory}%
  \BibitemOpen
  \bibfield  {author} {\bibinfo {author} {\bibfnamefont {J.~N.}\ \bibnamefont
  {Onuchic}}\ and\ \bibinfo {author} {\bibfnamefont {P.~G.}\ \bibnamefont
  {Wolynes}},\ }\href@noop {} {\bibfield  {journal} {\bibinfo  {journal} {Curr.
  Opin. Struct. Biol.}\ }\textbf {\bibinfo {volume} {14}},\ \bibinfo {pages}
  {70} (\bibinfo {year} {2004})}\BibitemShut {NoStop}%
\bibitem [{\citenamefont {Berezhkovskii}\ and\ \citenamefont
  {Szabo}(2005)}]{berezhkovskii2005one}%
  \BibitemOpen
  \bibfield  {author} {\bibinfo {author} {\bibfnamefont {A.}~\bibnamefont
  {Berezhkovskii}}\ and\ \bibinfo {author} {\bibfnamefont {A.}~\bibnamefont
  {Szabo}},\ }\href@noop {} {\bibfield  {journal} {\bibinfo  {journal} {J.
  Chem. Phys.}\ }\textbf {\bibinfo {volume} {122}},\ \bibinfo {pages} {014503}
  (\bibinfo {year} {2005})}\BibitemShut {NoStop}%
\bibitem [{\citenamefont {Best}\ and\ \citenamefont
  {Hummer}(2006)}]{best2006diffusive}%
  \BibitemOpen
  \bibfield  {author} {\bibinfo {author} {\bibfnamefont {R.~B.}\ \bibnamefont
  {Best}}\ and\ \bibinfo {author} {\bibfnamefont {G.}~\bibnamefont {Hummer}},\
  }\href@noop {} {\bibfield  {journal} {\bibinfo  {journal} {Phys. Rev. Lett.}\
  }\textbf {\bibinfo {volume} {96}},\ \bibinfo {pages} {228104} (\bibinfo
  {year} {2006})}\BibitemShut {NoStop}%
\bibitem [{\citenamefont {Zhang}\ \emph {et~al.}(2007)\citenamefont {Zhang},
  \citenamefont {Jasnow},\ and\ \citenamefont
  {Zuckerman}}]{zhang2007transition}%
  \BibitemOpen
  \bibfield  {author} {\bibinfo {author} {\bibfnamefont {B.~W.}\ \bibnamefont
  {Zhang}}, \bibinfo {author} {\bibfnamefont {D.}~\bibnamefont {Jasnow}}, \
  and\ \bibinfo {author} {\bibfnamefont {D.~M.}\ \bibnamefont {Zuckerman}},\
  }\href@noop {} {\bibfield  {journal} {\bibinfo  {journal} {J. Chem. Phys.}\
  }\textbf {\bibinfo {volume} {126}},\ \bibinfo {pages} {074504} (\bibinfo
  {year} {2007})}\BibitemShut {NoStop}%
\bibitem [{\citenamefont {Best}\ and\ \citenamefont
  {Hummer}(2011)}]{best2011diffusion}%
  \BibitemOpen
  \bibfield  {author} {\bibinfo {author} {\bibfnamefont {R.~B.}\ \bibnamefont
  {Best}}\ and\ \bibinfo {author} {\bibfnamefont {G.}~\bibnamefont {Hummer}},\
  }\href@noop {} {\bibfield  {journal} {\bibinfo  {journal} {Phys. Chem. Chem.
  Phys.}\ }\textbf {\bibinfo {volume} {13}},\ \bibinfo {pages} {16902}
  (\bibinfo {year} {2011})}\BibitemShut {NoStop}%
\bibitem [{\citenamefont {Hummer}(2005)}]{hummer2005position}%
  \BibitemOpen
  \bibfield  {author} {\bibinfo {author} {\bibfnamefont {G.}~\bibnamefont
  {Hummer}},\ }\href@noop {} {\bibfield  {journal} {\bibinfo  {journal} {New J.
  Phys.}\ }\textbf {\bibinfo {volume} {7}},\ \bibinfo {pages} {34} (\bibinfo
  {year} {2005})}\BibitemShut {NoStop}%
\bibitem [{\citenamefont {Berezhkovskii}\ and\ \citenamefont
  {Szabo}(2011)}]{berezhkovskii2011time}%
  \BibitemOpen
  \bibfield  {author} {\bibinfo {author} {\bibfnamefont {A.}~\bibnamefont
  {Berezhkovskii}}\ and\ \bibinfo {author} {\bibfnamefont {A.}~\bibnamefont
  {Szabo}},\ }\href@noop {} {\bibfield  {journal} {\bibinfo  {journal} {J.
  Chem. Phys.}\ }\textbf {\bibinfo {volume} {135}},\ \bibinfo {pages} {074108}
  (\bibinfo {year} {2011})}\BibitemShut {NoStop}%
\bibitem [{\citenamefont {Best}\ and\ \citenamefont
  {Hummer}(2005)}]{best2005reaction}%
  \BibitemOpen
  \bibfield  {author} {\bibinfo {author} {\bibfnamefont {R.~B.}\ \bibnamefont
  {Best}}\ and\ \bibinfo {author} {\bibfnamefont {G.}~\bibnamefont {Hummer}},\
  }\href@noop {} {\bibfield  {journal} {\bibinfo  {journal} {Proc. Natl. Acad.
  Sci. U.S.A.}\ }\textbf {\bibinfo {volume} {102}},\ \bibinfo {pages} {6732}
  (\bibinfo {year} {2005})}\BibitemShut {NoStop}%
\bibitem [{\citenamefont {Hinczewski}\ \emph {et~al.}(2010)\citenamefont
  {Hinczewski}, \citenamefont {von Hansen}, \citenamefont {Dzubiella},\ and\
  \citenamefont {Netz}}]{hinczewski2010diffusivity}%
  \BibitemOpen
  \bibfield  {author} {\bibinfo {author} {\bibfnamefont {M.}~\bibnamefont
  {Hinczewski}}, \bibinfo {author} {\bibfnamefont {Y.}~\bibnamefont {von
  Hansen}}, \bibinfo {author} {\bibfnamefont {J.}~\bibnamefont {Dzubiella}}, \
  and\ \bibinfo {author} {\bibfnamefont {R.~R.}\ \bibnamefont {Netz}},\
  }\href@noop {} {\bibfield  {journal} {\bibinfo  {journal} {J. Chem. Phys.}\
  }\textbf {\bibinfo {volume} {132}},\ \bibinfo {pages} {245103} (\bibinfo
  {year} {2010})}\BibitemShut {NoStop}%
\bibitem [{\citenamefont {Mugnai}\ and\ \citenamefont
  {Elber}(2015)}]{mugnai2015extracting}%
  \BibitemOpen
  \bibfield  {author} {\bibinfo {author} {\bibfnamefont {M.~L.}\ \bibnamefont
  {Mugnai}}\ and\ \bibinfo {author} {\bibfnamefont {R.}~\bibnamefont {Elber}},\
  }\href@noop {} {\bibfield  {journal} {\bibinfo  {journal} {J. Chem. Phys.}\
  }\textbf {\bibinfo {volume} {142}},\ \bibinfo {pages} {014105} (\bibinfo
  {year} {2015})}\BibitemShut {NoStop}%
\bibitem [{\citenamefont {Best}\ and\ \citenamefont
  {Hummer}(2010)}]{best2010coordinate}%
  \BibitemOpen
  \bibfield  {author} {\bibinfo {author} {\bibfnamefont {R.~B.}\ \bibnamefont
  {Best}}\ and\ \bibinfo {author} {\bibfnamefont {G.}~\bibnamefont {Hummer}},\
  }\href@noop {} {\bibfield  {journal} {\bibinfo  {journal} {Proc. Natl. Acad.
  Sci. U.S.A.}\ }\textbf {\bibinfo {volume} {107}},\ \bibinfo {pages} {1088}
  (\bibinfo {year} {2010})}\BibitemShut {NoStop}%
\bibitem [{\citenamefont {Tiwary}\ and\ \citenamefont
  {Berne}(2017)}]{tiwary2017predicting}%
  \BibitemOpen
  \bibfield  {author} {\bibinfo {author} {\bibfnamefont {P.}~\bibnamefont
  {Tiwary}}\ and\ \bibinfo {author} {\bibfnamefont {B.}~\bibnamefont {Berne}},\
  }\href@noop {} {\bibfield  {journal} {\bibinfo  {journal} {J. Chem. Phys.}\
  }\textbf {\bibinfo {volume} {147}},\ \bibinfo {pages} {152701} (\bibinfo
  {year} {2017})}\BibitemShut {NoStop}%
\bibitem [{\citenamefont {Du}\ \emph {et~al.}(1998)\citenamefont {Du},
  \citenamefont {Pande}, \citenamefont {Grosberg}, \citenamefont {Tanaka},\
  and\ \citenamefont {Shakhnovich}}]{du1998transition}%
  \BibitemOpen
  \bibfield  {author} {\bibinfo {author} {\bibfnamefont {R.}~\bibnamefont
  {Du}}, \bibinfo {author} {\bibfnamefont {V.~S.}\ \bibnamefont {Pande}},
  \bibinfo {author} {\bibfnamefont {A.~Y.}\ \bibnamefont {Grosberg}}, \bibinfo
  {author} {\bibfnamefont {T.}~\bibnamefont {Tanaka}}, \ and\ \bibinfo {author}
  {\bibfnamefont {E.~I.}\ \bibnamefont {Shakhnovich}},\ }\href@noop {}
  {\bibfield  {journal} {\bibinfo  {journal} {J. Chem. Phys.}\ }\textbf
  {\bibinfo {volume} {108}},\ \bibinfo {pages} {334} (\bibinfo {year}
  {1998})}\BibitemShut {NoStop}%
\bibitem [{\citenamefont {Neupane}\ \emph {et~al.}(2016)\citenamefont
  {Neupane}, \citenamefont {Foster}, \citenamefont {Dee}, \citenamefont {Yu},
  \citenamefont {Wang},\ and\ \citenamefont {Woodside}}]{neupane2016direct}%
  \BibitemOpen
  \bibfield  {author} {\bibinfo {author} {\bibfnamefont {K.}~\bibnamefont
  {Neupane}}, \bibinfo {author} {\bibfnamefont {D.~A.~N.}\ \bibnamefont
  {Foster}}, \bibinfo {author} {\bibfnamefont {D.~R.}\ \bibnamefont {Dee}},
  \bibinfo {author} {\bibfnamefont {H.}~\bibnamefont {Yu}}, \bibinfo {author}
  {\bibfnamefont {F.}~\bibnamefont {Wang}}, \ and\ \bibinfo {author}
  {\bibfnamefont {M.~T.}\ \bibnamefont {Woodside}},\ }\href@noop {} {\bibfield
  {journal} {\bibinfo  {journal} {Science}\ }\textbf {\bibinfo {volume}
  {352}},\ \bibinfo {pages} {239} (\bibinfo {year} {2016})}\BibitemShut
  {NoStop}%
\bibitem [{\citenamefont {Makarov}(2017)}]{makarov2017reconciling}%
  \BibitemOpen
  \bibfield  {author} {\bibinfo {author} {\bibfnamefont {D.~E.}\ \bibnamefont
  {Makarov}},\ }\href@noop {} {\bibfield  {journal} {\bibinfo  {journal} {J.
  Chem. Phys.}\ }\textbf {\bibinfo {volume} {146}},\ \bibinfo {pages} {071101}
  (\bibinfo {year} {2017})}\BibitemShut {NoStop}%
\bibitem [{\citenamefont {Satija}\ \emph {et~al.}(2017)\citenamefont {Satija},
  \citenamefont {Das},\ and\ \citenamefont {Makarov}}]{satija2017transition}%
  \BibitemOpen
  \bibfield  {author} {\bibinfo {author} {\bibfnamefont {R.}~\bibnamefont
  {Satija}}, \bibinfo {author} {\bibfnamefont {A.}~\bibnamefont {Das}}, \ and\
  \bibinfo {author} {\bibfnamefont {D.~E.}\ \bibnamefont {Makarov}},\
  }\href@noop {} {\bibfield  {journal} {\bibinfo  {journal} {J. Chem. Phys.}\
  }\textbf {\bibinfo {volume} {147}},\ \bibinfo {pages} {152707} (\bibinfo
  {year} {2017})}\BibitemShut {NoStop}%
\bibitem [{\citenamefont {Jacobs}\ and\ \citenamefont
  {Shakhnovich}(2016)}]{jacobs2016structure}%
  \BibitemOpen
  \bibfield  {author} {\bibinfo {author} {\bibfnamefont {W.~M.}\ \bibnamefont
  {Jacobs}}\ and\ \bibinfo {author} {\bibfnamefont {E.~I.}\ \bibnamefont
  {Shakhnovich}},\ }\href@noop {} {\bibfield  {journal} {\bibinfo  {journal}
  {Biophys. J.}\ }\textbf {\bibinfo {volume} {111}},\ \bibinfo {pages} {925}
  (\bibinfo {year} {2016})}\BibitemShut {NoStop}%
\bibitem [{\citenamefont {Mu{\~n}oz}\ and\ \citenamefont
  {Eaton}(1999)}]{munoz1999simple}%
  \BibitemOpen
  \bibfield  {author} {\bibinfo {author} {\bibfnamefont {V.}~\bibnamefont
  {Mu{\~n}oz}}\ and\ \bibinfo {author} {\bibfnamefont {W.~A.}\ \bibnamefont
  {Eaton}},\ }\href@noop {} {\bibfield  {journal} {\bibinfo  {journal} {Proc.
  Natl. Acad. Sci. U.S.A.}\ }\textbf {\bibinfo {volume} {96}},\ \bibinfo
  {pages} {11311} (\bibinfo {year} {1999})}\BibitemShut {NoStop}%
\bibitem [{\citenamefont {Alm}\ and\ \citenamefont
  {Baker}(1999)}]{alm1999prediction}%
  \BibitemOpen
  \bibfield  {author} {\bibinfo {author} {\bibfnamefont {E.}~\bibnamefont
  {Alm}}\ and\ \bibinfo {author} {\bibfnamefont {D.}~\bibnamefont {Baker}},\
  }\href@noop {} {\bibfield  {journal} {\bibinfo  {journal} {Proc. Natl. Acad.
  Sci. U.S.A.}\ }\textbf {\bibinfo {volume} {96}},\ \bibinfo {pages} {11305}
  (\bibinfo {year} {1999})}\BibitemShut {NoStop}%
\bibitem [{\citenamefont {Galzitskaya}\ and\ \citenamefont
  {Finkelstein}(1999)}]{galzitskaya1999theoretical}%
  \BibitemOpen
  \bibfield  {author} {\bibinfo {author} {\bibfnamefont {O.~V.}\ \bibnamefont
  {Galzitskaya}}\ and\ \bibinfo {author} {\bibfnamefont {A.~V.}\ \bibnamefont
  {Finkelstein}},\ }\href@noop {} {\bibfield  {journal} {\bibinfo  {journal}
  {Proc. Natl. Acad. Sci. U.S.A.}\ }\textbf {\bibinfo {volume} {96}},\ \bibinfo
  {pages} {11299} (\bibinfo {year} {1999})}\BibitemShut {NoStop}%
\bibitem [{\citenamefont {Frenkel}(2016)}]{frenkel2016folding}%
  \BibitemOpen
  \bibfield  {author} {\bibinfo {author} {\bibfnamefont {D.}~\bibnamefont
  {Frenkel}},\ }\href@noop {} {\bibfield  {journal} {\bibinfo  {journal}
  {Biophys. J.}\ }\textbf {\bibinfo {volume} {111}},\ \bibinfo {pages} {893}
  (\bibinfo {year} {2016})}\BibitemShut {NoStop}%
\bibitem [{\citenamefont {Piana}\ \emph {et~al.}(2013)\citenamefont {Piana},
  \citenamefont {Lindorff-Larsen},\ and\ \citenamefont
  {Shaw}}]{piana2013atomic}%
  \BibitemOpen
  \bibfield  {author} {\bibinfo {author} {\bibfnamefont {S.}~\bibnamefont
  {Piana}}, \bibinfo {author} {\bibfnamefont {K.}~\bibnamefont
  {Lindorff-Larsen}}, \ and\ \bibinfo {author} {\bibfnamefont {D.~E.}\
  \bibnamefont {Shaw}},\ }\href@noop {} {\bibfield  {journal} {\bibinfo
  {journal} {Proc. Natl. Acad. Sci. U.S.A.}\ }\textbf {\bibinfo {volume}
  {110}},\ \bibinfo {pages} {5915} (\bibinfo {year} {2013})}\BibitemShut
  {NoStop}%
\bibitem [{\citenamefont {Thompson}\ \emph {et~al.}(1997)\citenamefont
  {Thompson}, \citenamefont {Eaton},\ and\ \citenamefont
  {Hofrichter}}]{thompson1997laser}%
  \BibitemOpen
  \bibfield  {author} {\bibinfo {author} {\bibfnamefont {P.~A.}\ \bibnamefont
  {Thompson}}, \bibinfo {author} {\bibfnamefont {W.~A.}\ \bibnamefont {Eaton}},
  \ and\ \bibinfo {author} {\bibfnamefont {J.}~\bibnamefont {Hofrichter}},\
  }\href@noop {} {\bibfield  {journal} {\bibinfo  {journal} {Biochemistry}\
  }\textbf {\bibinfo {volume} {36}},\ \bibinfo {pages} {9200} (\bibinfo {year}
  {1997})}\BibitemShut {NoStop}%
\bibitem [{\citenamefont {Lapidus}\ \emph {et~al.}(2000)\citenamefont
  {Lapidus}, \citenamefont {Eaton},\ and\ \citenamefont
  {Hofrichter}}]{lapidus2000measuring}%
  \BibitemOpen
  \bibfield  {author} {\bibinfo {author} {\bibfnamefont {L.~J.}\ \bibnamefont
  {Lapidus}}, \bibinfo {author} {\bibfnamefont {W.~A.}\ \bibnamefont {Eaton}},
  \ and\ \bibinfo {author} {\bibfnamefont {J.}~\bibnamefont {Hofrichter}},\
  }\href@noop {} {\bibfield  {journal} {\bibinfo  {journal} {Proc. Natl. Acad.
  Sci. U.S.A.}\ }\textbf {\bibinfo {volume} {97}},\ \bibinfo {pages} {7220}
  (\bibinfo {year} {2000})}\BibitemShut {NoStop}%
\bibitem [{\citenamefont {Garbuzynskiy}\ \emph {et~al.}(2013)\citenamefont
  {Garbuzynskiy}, \citenamefont {Ivankov}, \citenamefont {Bogatyreva},\ and\
  \citenamefont {Finkelstein}}]{garbuzynskiy2013golden}%
  \BibitemOpen
  \bibfield  {author} {\bibinfo {author} {\bibfnamefont {S.~O.}\ \bibnamefont
  {Garbuzynskiy}}, \bibinfo {author} {\bibfnamefont {D.~N.}\ \bibnamefont
  {Ivankov}}, \bibinfo {author} {\bibfnamefont {N.~S.}\ \bibnamefont
  {Bogatyreva}}, \ and\ \bibinfo {author} {\bibfnamefont {A.~V.}\ \bibnamefont
  {Finkelstein}},\ }\href@noop {} {\bibfield  {journal} {\bibinfo  {journal}
  {Proc. Natl. Acad. Sci. U.S.A.}\ }\textbf {\bibinfo {volume} {110}},\
  \bibinfo {pages} {147} (\bibinfo {year} {2013})}\BibitemShut {NoStop}%
\bibitem [{\citenamefont {Zimm}\ and\ \citenamefont
  {Bragg}(1959)}]{zimm1959theory}%
  \BibitemOpen
  \bibfield  {author} {\bibinfo {author} {\bibfnamefont {B.~H.}\ \bibnamefont
  {Zimm}}\ and\ \bibinfo {author} {\bibfnamefont {J.~K.}\ \bibnamefont
  {Bragg}},\ }\href@noop {} {\bibfield  {journal} {\bibinfo  {journal} {J.
  Chem. Phys.}\ }\textbf {\bibinfo {volume} {31}},\ \bibinfo {pages} {526}
  (\bibinfo {year} {1959})}\BibitemShut {NoStop}%
\bibitem [{\citenamefont {Dill}\ \emph {et~al.}(1993)\citenamefont {Dill},
  \citenamefont {Fiebig},\ and\ \citenamefont {Chan}}]{dill1993cooperativity}%
  \BibitemOpen
  \bibfield  {author} {\bibinfo {author} {\bibfnamefont {K.~A.}\ \bibnamefont
  {Dill}}, \bibinfo {author} {\bibfnamefont {K.~M.}\ \bibnamefont {Fiebig}}, \
  and\ \bibinfo {author} {\bibfnamefont {H.~S.}\ \bibnamefont {Chan}},\
  }\href@noop {} {\bibfield  {journal} {\bibinfo  {journal} {Proc. Natl. Acad.
  Sci. U.S.A.}\ }\textbf {\bibinfo {volume} {90}},\ \bibinfo {pages} {1942}
  (\bibinfo {year} {1993})}\BibitemShut {NoStop}%
\bibitem [{\citenamefont {Kubelka}\ \emph {et~al.}(2008)\citenamefont
  {Kubelka}, \citenamefont {Henry}, \citenamefont {Cellmer}, \citenamefont
  {Hofrichter},\ and\ \citenamefont {Eaton}}]{kubelka2008chemical}%
  \BibitemOpen
  \bibfield  {author} {\bibinfo {author} {\bibfnamefont {J.}~\bibnamefont
  {Kubelka}}, \bibinfo {author} {\bibfnamefont {E.~R.}\ \bibnamefont {Henry}},
  \bibinfo {author} {\bibfnamefont {T.}~\bibnamefont {Cellmer}}, \bibinfo
  {author} {\bibfnamefont {J.}~\bibnamefont {Hofrichter}}, \ and\ \bibinfo
  {author} {\bibfnamefont {W.~A.}\ \bibnamefont {Eaton}},\ }\href@noop {}
  {\bibfield  {journal} {\bibinfo  {journal} {Proc. Natl. Acad. Sci. U.S.A.}\
  }\textbf {\bibinfo {volume} {105}},\ \bibinfo {pages} {18655} (\bibinfo
  {year} {2008})}\BibitemShut {NoStop}%
\bibitem [{\citenamefont {Henry}\ \emph {et~al.}(2013)\citenamefont {Henry},
  \citenamefont {Best},\ and\ \citenamefont {Eaton}}]{henry2013comparing}%
  \BibitemOpen
  \bibfield  {author} {\bibinfo {author} {\bibfnamefont {E.~R.}\ \bibnamefont
  {Henry}}, \bibinfo {author} {\bibfnamefont {R.~B.}\ \bibnamefont {Best}}, \
  and\ \bibinfo {author} {\bibfnamefont {W.~A.}\ \bibnamefont {Eaton}},\
  }\href@noop {} {\bibfield  {journal} {\bibinfo  {journal} {Proc. Natl. Acad.
  Sci. U.S.A.}\ }\textbf {\bibinfo {volume} {110}},\ \bibinfo {pages} {17880}
  (\bibinfo {year} {2013})}\BibitemShut {NoStop}%
\bibitem [{\citenamefont {Best}\ \emph {et~al.}(2013)\citenamefont {Best},
  \citenamefont {Hummer},\ and\ \citenamefont {Eaton}}]{best2013native}%
  \BibitemOpen
  \bibfield  {author} {\bibinfo {author} {\bibfnamefont {R.~B.}\ \bibnamefont
  {Best}}, \bibinfo {author} {\bibfnamefont {G.}~\bibnamefont {Hummer}}, \ and\
  \bibinfo {author} {\bibfnamefont {W.~A.}\ \bibnamefont {Eaton}},\ }\href@noop
  {} {\bibfield  {journal} {\bibinfo  {journal} {Proc. Natl. Acad. Sci.
  U.S.A.}\ }\textbf {\bibinfo {volume} {110}},\ \bibinfo {pages} {17874}
  (\bibinfo {year} {2013})}\BibitemShut {NoStop}%
\bibitem [{\citenamefont {Metzner}\ \emph {et~al.}(2009)\citenamefont
  {Metzner}, \citenamefont {Sch{\"u}tte},\ and\ \citenamefont
  {Vanden-Eijnden}}]{metzner2009transition}%
  \BibitemOpen
  \bibfield  {author} {\bibinfo {author} {\bibfnamefont {P.}~\bibnamefont
  {Metzner}}, \bibinfo {author} {\bibfnamefont {C.}~\bibnamefont
  {Sch{\"u}tte}}, \ and\ \bibinfo {author} {\bibfnamefont {E.}~\bibnamefont
  {Vanden-Eijnden}},\ }\href@noop {} {\bibfield  {journal} {\bibinfo  {journal}
  {Multiscale Model. Sim.}\ }\textbf {\bibinfo {volume} {7}},\ \bibinfo {pages}
  {1192} (\bibinfo {year} {2009})}\BibitemShut {NoStop}%
\bibitem [{\citenamefont {Metropolis}\ \emph {et~al.}(1953)\citenamefont
  {Metropolis}, \citenamefont {Rosenbluth}, \citenamefont {Rosenbluth},
  \citenamefont {Teller},\ and\ \citenamefont
  {Teller}}]{metropolis1953equation}%
  \BibitemOpen
  \bibfield  {author} {\bibinfo {author} {\bibfnamefont {N.}~\bibnamefont
  {Metropolis}}, \bibinfo {author} {\bibfnamefont {A.~W.}\ \bibnamefont
  {Rosenbluth}}, \bibinfo {author} {\bibfnamefont {M.~N.}\ \bibnamefont
  {Rosenbluth}}, \bibinfo {author} {\bibfnamefont {A.~H.}\ \bibnamefont
  {Teller}}, \ and\ \bibinfo {author} {\bibfnamefont {E.}~\bibnamefont
  {Teller}},\ }\href@noop {} {\bibfield  {journal} {\bibinfo  {journal} {J.
  Chem. Phys.}\ }\textbf {\bibinfo {volume} {21}},\ \bibinfo {pages} {1087}
  (\bibinfo {year} {1953})}\BibitemShut {NoStop}%
\bibitem [{\citenamefont {Krivov}\ and\ \citenamefont
  {Karplus}(2004)}]{krivov2004hidden}%
  \BibitemOpen
  \bibfield  {author} {\bibinfo {author} {\bibfnamefont {S.~V.}\ \bibnamefont
  {Krivov}}\ and\ \bibinfo {author} {\bibfnamefont {M.}~\bibnamefont
  {Karplus}},\ }\href@noop {} {\bibfield  {journal} {\bibinfo  {journal} {Proc.
  Natl. Acad. Sci. U.S.A.}\ }\textbf {\bibinfo {volume} {101}},\ \bibinfo
  {pages} {14766} (\bibinfo {year} {2004})}\BibitemShut {NoStop}%
\bibitem [{\citenamefont {Krivov}\ and\ \citenamefont
  {Karplus}(2006)}]{krivov2006one}%
  \BibitemOpen
  \bibfield  {author} {\bibinfo {author} {\bibfnamefont {S.~V.}\ \bibnamefont
  {Krivov}}\ and\ \bibinfo {author} {\bibfnamefont {M.}~\bibnamefont
  {Karplus}},\ }\href@noop {} {\bibfield  {journal} {\bibinfo  {journal} {J.
  Phys. Chem. B}\ }\textbf {\bibinfo {volume} {110}},\ \bibinfo {pages} {12689}
  (\bibinfo {year} {2006})}\BibitemShut {NoStop}%
\bibitem [{\citenamefont {Ding}\ \emph {et~al.}(2002)\citenamefont {Ding},
  \citenamefont {Dokholyan}, \citenamefont {Buldyrev}, \citenamefont
  {Stanley},\ and\ \citenamefont {Shakhnovich}}]{ding2002direct}%
  \BibitemOpen
  \bibfield  {author} {\bibinfo {author} {\bibfnamefont {F.}~\bibnamefont
  {Ding}}, \bibinfo {author} {\bibfnamefont {N.~V.}\ \bibnamefont {Dokholyan}},
  \bibinfo {author} {\bibfnamefont {S.~V.}\ \bibnamefont {Buldyrev}}, \bibinfo
  {author} {\bibfnamefont {H.~E.}\ \bibnamefont {Stanley}}, \ and\ \bibinfo
  {author} {\bibfnamefont {E.~I.}\ \bibnamefont {Shakhnovich}},\ }\href@noop {}
  {\bibfield  {journal} {\bibinfo  {journal} {Biophys. J.}\ }\textbf {\bibinfo
  {volume} {83}},\ \bibinfo {pages} {3525} (\bibinfo {year}
  {2002})}\BibitemShut {NoStop}%
\bibitem [{\citenamefont {Chung}\ \emph {et~al.}(2009)\citenamefont {Chung},
  \citenamefont {Louis},\ and\ \citenamefont {Eaton}}]{chung2009experimental}%
  \BibitemOpen
  \bibfield  {author} {\bibinfo {author} {\bibfnamefont {H.~S.}\ \bibnamefont
  {Chung}}, \bibinfo {author} {\bibfnamefont {J.~M.}\ \bibnamefont {Louis}}, \
  and\ \bibinfo {author} {\bibfnamefont {W.~A.}\ \bibnamefont {Eaton}},\
  }\href@noop {} {\bibfield  {journal} {\bibinfo  {journal} {Proc. Natl. Acad.
  Sci. U.S.A.}\ }\textbf {\bibinfo {volume} {106}},\ \bibinfo {pages} {11837}
  (\bibinfo {year} {2009})}\BibitemShut {NoStop}%
\bibitem [{\citenamefont {Gillespie}(1977)}]{gillespie1977exact}%
  \BibitemOpen
  \bibfield  {author} {\bibinfo {author} {\bibfnamefont {D.~T.}\ \bibnamefont
  {Gillespie}},\ }\href@noop {} {\bibfield  {journal} {\bibinfo  {journal} {J.
  Phys. Chem.}\ }\textbf {\bibinfo {volume} {81}},\ \bibinfo {pages} {2340}
  (\bibinfo {year} {1977})}\BibitemShut {NoStop}%
\bibitem [{\citenamefont {Chaudhury}\ and\ \citenamefont
  {Makarov}(2010)}]{chaudhury2010harmonic}%
  \BibitemOpen
  \bibfield  {author} {\bibinfo {author} {\bibfnamefont {S.}~\bibnamefont
  {Chaudhury}}\ and\ \bibinfo {author} {\bibfnamefont {D.~E.}\ \bibnamefont
  {Makarov}},\ }\href@noop {} {\bibfield  {journal} {\bibinfo  {journal} {J.
  Chem. Phys.}\ }\textbf {\bibinfo {volume} {133}},\ \bibinfo {pages} {034118}
  (\bibinfo {year} {2010})}\BibitemShut {NoStop}%
\bibitem [{\citenamefont {Krivov}(2010)}]{krivov2010protein}%
  \BibitemOpen
  \bibfield  {author} {\bibinfo {author} {\bibfnamefont {S.~V.}\ \bibnamefont
  {Krivov}},\ }\href@noop {} {\bibfield  {journal} {\bibinfo  {journal} {PLoS
  Comp. Biol.}\ }\textbf {\bibinfo {volume} {6}},\ \bibinfo {pages} {e1000921}
  (\bibinfo {year} {2010})}\BibitemShut {NoStop}%
\bibitem [{\citenamefont {Forney}(1973)}]{forney1973viterbi}%
  \BibitemOpen
  \bibfield  {author} {\bibinfo {author} {\bibfnamefont {G.~D.}\ \bibnamefont
  {Forney}},\ }\href@noop {} {\bibfield  {journal} {\bibinfo  {journal} {Proc.
  IEEE}\ }\textbf {\bibinfo {volume} {61}},\ \bibinfo {pages} {268} (\bibinfo
  {year} {1973})}\BibitemShut {NoStop}%
\bibitem [{hmm()}]{hmmlearn}%
  \BibitemOpen
  \href@noop {} {\enquote {\bibinfo {title} {hmmlearn --- hmmlearn 0.2.1
  documentation},}\ }\bibinfo {howpublished}
  {\url{http://hmmlearn.readthedocs.io/en/latest/}},\ \bibinfo {note}
  {accessed: June 19, 2018}\BibitemShut {NoStop}%
\bibitem [{\citenamefont {Bowman}\ \emph {et~al.}(2014)\citenamefont {Bowman},
  \citenamefont {Pande},\ and\ \citenamefont
  {No{\'e}}}]{bowman2014introduction}%
  \BibitemOpen
  \bibfield  {author} {\bibinfo {author} {\bibfnamefont {G.~R.}\ \bibnamefont
  {Bowman}}, \bibinfo {author} {\bibfnamefont {V.~S.}\ \bibnamefont {Pande}}, \
  and\ \bibinfo {author} {\bibfnamefont {F.}~\bibnamefont {No{\'e}}},\ }in\
  \href@noop {} {\emph {\bibinfo {booktitle} {An introduction to Markov state
  models and their application to long timescale molecular simulation}}}\
  (\bibinfo  {publisher} {Springer},\ \bibinfo {year} {2014})\ pp.\ \bibinfo
  {pages} {1--6}\BibitemShut {NoStop}%
\bibitem [{\citenamefont {Chodera}\ and\ \citenamefont
  {No{\'e}}(2014)}]{chodera2014markov}%
  \BibitemOpen
  \bibfield  {author} {\bibinfo {author} {\bibfnamefont {J.~D.}\ \bibnamefont
  {Chodera}}\ and\ \bibinfo {author} {\bibfnamefont {F.}~\bibnamefont
  {No{\'e}}},\ }\href@noop {} {\bibfield  {journal} {\bibinfo  {journal} {Curr.
  Opin. Struct. Biol.}\ }\textbf {\bibinfo {volume} {25}},\ \bibinfo {pages}
  {135} (\bibinfo {year} {2014})}\BibitemShut {NoStop}%
\bibitem [{\citenamefont {Lindorff-Larsen}\ \emph {et~al.}(2011)\citenamefont
  {Lindorff-Larsen}, \citenamefont {Piana}, \citenamefont {Dror},\ and\
  \citenamefont {Shaw}}]{lindorff2011fast}%
  \BibitemOpen
  \bibfield  {author} {\bibinfo {author} {\bibfnamefont {K.}~\bibnamefont
  {Lindorff-Larsen}}, \bibinfo {author} {\bibfnamefont {S.}~\bibnamefont
  {Piana}}, \bibinfo {author} {\bibfnamefont {R.~O.}\ \bibnamefont {Dror}}, \
  and\ \bibinfo {author} {\bibfnamefont {D.~E.}\ \bibnamefont {Shaw}},\
  }\href@noop {} {\bibfield  {journal} {\bibinfo  {journal} {Science}\ }\textbf
  {\bibinfo {volume} {334}},\ \bibinfo {pages} {517} (\bibinfo {year}
  {2011})}\BibitemShut {NoStop}%
\bibitem [{\citenamefont {Ptitsyn}(1973)}]{ptitsyn1973stages}%
  \BibitemOpen
  \bibfield  {author} {\bibinfo {author} {\bibfnamefont {O.}~\bibnamefont
  {Ptitsyn}},\ }\href@noop {} {\bibfield  {journal} {\bibinfo  {journal}
  {Doklady Akademii Nauk SSSR}\ }\textbf {\bibinfo {volume} {210}},\ \bibinfo
  {pages} {1213} (\bibinfo {year} {1973})}\BibitemShut {NoStop}%
\bibitem [{\citenamefont {Maity}\ \emph {et~al.}(2005)\citenamefont {Maity},
  \citenamefont {Maity}, \citenamefont {Krishna}, \citenamefont {Mayne},\ and\
  \citenamefont {Englander}}]{maity2005protein}%
  \BibitemOpen
  \bibfield  {author} {\bibinfo {author} {\bibfnamefont {H.}~\bibnamefont
  {Maity}}, \bibinfo {author} {\bibfnamefont {M.}~\bibnamefont {Maity}},
  \bibinfo {author} {\bibfnamefont {M.~M.}\ \bibnamefont {Krishna}}, \bibinfo
  {author} {\bibfnamefont {L.}~\bibnamefont {Mayne}}, \ and\ \bibinfo {author}
  {\bibfnamefont {S.~W.}\ \bibnamefont {Englander}},\ }\href@noop {} {\bibfield
   {journal} {\bibinfo  {journal} {Proc. Natl. Acad. Sci. U.S.A.}\ }\textbf
  {\bibinfo {volume} {102}},\ \bibinfo {pages} {4741} (\bibinfo {year}
  {2005})}\BibitemShut {NoStop}%
\bibitem [{\citenamefont {Rollins}\ and\ \citenamefont
  {Dill}(2014)}]{rollins2014general}%
  \BibitemOpen
  \bibfield  {author} {\bibinfo {author} {\bibfnamefont {G.~C.}\ \bibnamefont
  {Rollins}}\ and\ \bibinfo {author} {\bibfnamefont {K.~A.}\ \bibnamefont
  {Dill}},\ }\href@noop {} {\bibfield  {journal} {\bibinfo  {journal} {J. Am.
  Chem. Soc.}\ }\textbf {\bibinfo {volume} {136}},\ \bibinfo {pages} {11420}
  (\bibinfo {year} {2014})}\BibitemShut {NoStop}%
\bibitem [{\citenamefont {Adhikari}\ \emph {et~al.}(2012)\citenamefont
  {Adhikari}, \citenamefont {Freed},\ and\ \citenamefont
  {Sosnick}}]{adhikari2012novo}%
  \BibitemOpen
  \bibfield  {author} {\bibinfo {author} {\bibfnamefont {A.~N.}\ \bibnamefont
  {Adhikari}}, \bibinfo {author} {\bibfnamefont {K.~F.}\ \bibnamefont {Freed}},
  \ and\ \bibinfo {author} {\bibfnamefont {T.~R.}\ \bibnamefont {Sosnick}},\
  }\href@noop {} {\bibfield  {journal} {\bibinfo  {journal} {Proc. Natl. Acad.
  Sci. U.S.A.}\ }\textbf {\bibinfo {volume} {109}},\ \bibinfo {pages} {17442}
  (\bibinfo {year} {2012})}\BibitemShut {NoStop}%
\bibitem [{\citenamefont {Sgouralis}\ \emph {et~al.}(2018)\citenamefont
  {Sgouralis}, \citenamefont {Whitmore}, \citenamefont {Lapidus}, \citenamefont
  {Comstock},\ and\ \citenamefont {Press{\'e}}}]{sgouralis2018single}%
  \BibitemOpen
  \bibfield  {author} {\bibinfo {author} {\bibfnamefont {I.}~\bibnamefont
  {Sgouralis}}, \bibinfo {author} {\bibfnamefont {M.}~\bibnamefont {Whitmore}},
  \bibinfo {author} {\bibfnamefont {L.}~\bibnamefont {Lapidus}}, \bibinfo
  {author} {\bibfnamefont {M.~J.}\ \bibnamefont {Comstock}}, \ and\ \bibinfo
  {author} {\bibfnamefont {S.}~\bibnamefont {Press{\'e}}},\ }\href@noop {}
  {\bibfield  {journal} {\bibinfo  {journal} {J. Chem. Phys.}\ }\textbf
  {\bibinfo {volume} {148}},\ \bibinfo {pages} {123320} (\bibinfo {year}
  {2018})}\BibitemShut {NoStop}%
\bibitem [{\citenamefont {Sgouralis}\ and\ \citenamefont
  {Press{\'e}}(2017)}]{sgouralis2017icon}%
  \BibitemOpen
  \bibfield  {author} {\bibinfo {author} {\bibfnamefont {I.}~\bibnamefont
  {Sgouralis}}\ and\ \bibinfo {author} {\bibfnamefont {S.}~\bibnamefont
  {Press{\'e}}},\ }\href@noop {} {\bibfield  {journal} {\bibinfo  {journal}
  {Biophys. J.}\ }\textbf {\bibinfo {volume} {112}},\ \bibinfo {pages} {2117}
  (\bibinfo {year} {2017})}\BibitemShut {NoStop}%
\end{thebibliography}
\end{document}